\documentclass[reprint,amsmath,amssymb,aps,prm]{revtex4-2}
\usepackage{graphicx}
\usepackage{dcolumn}
\usepackage{bm}
\usepackage{courier}
\usepackage{amsmath,amssymb}
\usepackage{mathrsfs}
\usepackage{bm}
\usepackage{supertabular}
\usepackage{float}
\usepackage{soul}
\usepackage{color}
\usepackage{natbib}
\usepackage{placeins}
\usepackage{array}
\usepackage{multirow}
\usepackage{afterpage}
\usepackage{lipsum}
\usepackage[utf8]{inputenc}
\usepackage{csquotes}
\usepackage{braket}
\usepackage{booktabs}
\usepackage{epstopdf}
\usepackage{gensymb}
\begin{document}
\preprint{APS/123-QED}
\title{Tuning Nonradiative Recombination via Cation Substitution in Inorganic Antiperovskite Nitrides}

\author{Sanchi Monga}
 \email{sanchi@physics.iitd.ac.in[SM]}
\author{Saswata Bhattacharya}
\email{saswata@physics.iitd.ac.in [SB]}
\affiliation{Department of Physics, Indian Institute of Technology Delhi, New Delhi 110016, India}

\begin{abstract}
Inorganic antiperovskite nitrides have recently emerged as promising materials for photovoltaic applications, yet their nonradiative recombination dynamics remain largely unexplored. Here, we examine the influence of X-site cation substitution on the nonradiative electron–hole recombination in $\mathrm{X_3NSb}$ (X = Ca, Sr, and Ba). Ca- and Sr-based compounds adopt a cubic phase, whereas Ba stabilizes in a hexagonal structure, introducing pronounced symmetry-driven effects. To separate symmetry effects from cation chemistry, we also examine the hexagonal polymorph of Sr$_3$NSb ($\mathrm{Sr_3NSb_{hexa}}$). Substituting Ca with Sr narrows the band gap, suppresses octahedral and band-edge fluctuations, reduces nonadiabatic (NA) coupling by $\sim$54$\%$, and extends carrier lifetimes by a factor of 2.5. In $\mathrm{Sr_3NSb_{hexa}}$, the combination of larger band gap and enhanced band gap fluctuations$-$leading to faster dephasing$-$further slows down recombination by 41$\%$. In contrast, in $\mathrm{Ba_3NSb_{hexa}}$, enhanced NA coupling accelerates recombination relative to $\mathrm{Sr_3NSb_{hexa}}$. Overall, recombination lifetimes are dictated by the interplay between band gap, NA coupling strength, and decoherence time, with $\mathrm{Sr_3NSb_{hexa}}$ exhibiting the longest lifetime. These findings highlight the coupled influence of cation chemistry and crystal symmetry in tailoring carrier dynamics for high-performance antiperovskite-based optoelectronics materials.
\end{abstract}
\maketitle

\section{Introduction}
Organic-inorganic lead halide perovskites (LHPs), such as $\mathrm{CH_3NH_3PbI_3}$ ($\mathrm{MAPbI_3}$), have emerged as promising photovoltaic materials owing to their favorable optoelectronic properties, including suitable band gaps, strong optical absorption, long carrier diffusion lengths, and extended carrier lifetimes \cite{Green2014,Yin2015,kim2020high,yu2020miscellaneous}. These attributes have driven a rapid increase in photo-conversion efficiency (PCE), rising from 3.8$\%$ to over 26$\%$ in recent years \cite{nrel}. Despite this remarkable progress, the commercial viability of LHPs remains hindered by their environmental instability and lead toxicity~\cite{schileo2021lead, ju2018toward}. These limitations motivate an intensive search for lead-free compositions that can simultaneously offer enhanced stability and desirable optoelectronic properties. 

Antiperovskites, derived from an ionic inversion of the conventional perovskite lattice, have recently attracted growing attention as promising candidates for photovoltaic and optoelectronic applications \cite{chi2002new,beznosikov2003predicted,heinselman2019thin,mochizuki2020theoretical,zhong2021structure,wang2020antiperovskites,dai2019bi,kang2022antiperovskite,gabler2004sr3n}. These compounds adopt the general formula X$_3$BA, where X is a cation and A and B are anions of different sizes. The inverted ionic configuration fundamentally modifies electronic structure and lattice dynamics relative to LHPs, while also substantially expanding the accessible compositional and structural space. However, antiperovskites remain comparatively underexplored, and to date only Mg$_3$NSb has been experimentally synthesized as a thin film \cite{heinselman2019thin}.

In our previous work \cite{monga2024theoretical}, we carried out a systematic first-principles investigation of the electronic, optical, excitonic, and polaronic properties of inorganic antiperovskite nitrides $\mathrm{X_3NA}$ (X = Mg, Ca, Sr, Ba; A = As, Sb). Within this family, Ca$_3$NSb, Sr$_3$NSb, and Ba$_3$NSb emerged as particularly promising direct-band-gap photovoltaic candidates, exhibiting optimal band gaps, low electron and hole effective masses ($\frac{m_{e/h}^*}{m_e}$ $<$ 1), small exciton binding energies ($<$ 30 meV), and high carrier mobilities exceeding those of the benchmark MAPbI$_3$ perovskite \cite{Frost2017}. Despite these favorable properties, their nonradiative recombination dynamics have not yet been explored.

Photovoltaic performance is critically limited by nonradiative recombination processes, which dissipate photoexcited carrier energy as heat. In recent years, \textit{ab initio} nonadiabatic molecular dynamics (NAMD) simulations have substantially advanced the understanding of carrier lifetimes in LHPs, demonstrating that nonradiative pathways can be strategically suppressed through strategies such as surface passivation, compositional engineering, and dimensionality control \cite{wang2025self,wang2024detrimental,panigrahi2022tailoring,du2020crystal,chen2019cation,de2018efficient,nayak2025optimizing}. Several of these computational predictions have also been corroborated experimentally \cite{wang2025self,panigrahi2022tailoring,wang2024detrimental,du2020crystal}.

In this work, we investigate the nonradiative recombination dynamics of X$_3$NSb (X = Ca, Sr, Ba) antiperovskites using real-time time-dependent density functional theory (TDDFT) \cite{runge1984density} combined with NAMD. These compounds span two distinct crystallographic motifs. Ca$_3$NSb and Sr$_3$NSb crystallize in the high-symmetry cubic $\mathit{Pm\bar{3}m}$ phase, enabling a controlled assessment of alkaline-earth cation size effects at fixed symmetry. In contrast, Ba$_3$NSb stabilizes in a lower-symmetry hexagonal $\mathit{P6_3/mmc}$ structure. To disentangle symmetry-driven effects from purely chemical ones, we additionally consider the hexagonal $\mathit{P6_3/mmc}$ polymorph of Sr$_3$NSb, allowing a direct comparison between cubic and hexagonal phases at identical chemical composition. Hexagonal phases are denoted by the subscript ``hexa" (e.g., $\mathrm{Ba_3NSb_{hexa}}$ and $\mathrm{Sr_3NSb_{hexa}}$), while compositions without subscripts refer exclusively to the cubic $\mathit{Pm\bar{3}m}$ phase. Together, this four-model set constitutes a minimal yet physically well-justified framework for elucidating how cation chemistry and crystal symmetry independently and cooperatively govern lattice fluctuations, electronic structure, nonadiabatic couplings, and nonradiative recombination lifetimes in antiperovskite nitrides. Our results establish a clear structure-property-dynamics relationship and provide fundamental design principles for engineering antiperovskite materials with suppressed nonradiative losses.

\section{Computational Details}
We perform \textit{ab initio} NAMD simulations using the decoherence-induced surface hopping (DISH) method \cite{craig2005trajectory, jaeger2012decoherence}, within the classical path approximation, as implemented in the PYXAID package \cite{akimov2013pyxaid, akimov2014advanced}. In this framework, the nuclear motion—associated with heavier and slower atoms—is treated semiclassically, whereas the electronic degrees of freedom are described quantum mechanically using TDDFT \cite{runge1984density}. Decoherence, arising from elastic electron–phonon scattering that disrupts superpositions between electronic states, plays a critical role in regulating quantum transitions. Since decoherence typically occurs on timescales shorter than those of interband transitions, its inclusion is essential for accurately capturing nonradiative electron–hole recombination processes. In the DISH algorithm, quantum transitions emerge as a consequence of decoherence, which provides the physical basis for surface hops. Decoherence time is estimated as the pure-dephasing time, evaluated via the second-order cumulant approximation of optical response theory \cite{hamm2005principles}.

First-principles density functional theory (DFT) calculations \cite{hohenberg1964inhomogeneous,kohn1965self} are performed using the Vienna \textit{ab initio} Simulation Package (VASP) \cite{kresse1996efficiency,kresse1999ultrasoft}. The projector augmented-wave (PAW) method \cite{blochl1994projector} is employed to accurately describe the interactions between valence electrons and the ionic cores. Structural optimization is carried out using the generalized gradient approximation (GGA) with the Perdew-Burke-Ernzerhof (PBE) exchange-correlation functional \cite{perdew1996generalized}. A $\Gamma$-centered $\mathbf{k}$-point mesh of 6$\times$6$\times$6 and a plane-wave cutoff energy of 550 eV are used to ensure convergence. The atomic positions are relaxed until the Hellmann–Feynman forces on each atom are less than 10$^{-3}$ eV/$\text{\r{A}}$. The calculated lattice parameters are listed in Table 1 of the Supporting Information (SI) \cite{Supporting_Information} and they show good agreement with previously reported values \cite{stoiber2019perovskite,gabler2004sr3n}. For electronic structure calculations, the hybrid Heyd–Scuseria–Ernzerhof (HSE06) functional \cite{krukau2006influence} is employed, including the spin–orbit coupling (SOC).

\textit{Ab initio} molecular dynamics (AIMD) simulations are carried out using the Quantum Espresso (QE) package \cite{giannozzi2009quantum}. Supercell of dimension 2$\times$2$\times$2 are constructed for all studied antiperovskites. Following geometry optimization, these supercells are heated to 300 K for a total simulation time of 13 ps with a time step of 1 fs, employing the Verlet algorithm within the canonical (NVT) ensemble. The initial 8 ps of each trajectory is considered for thermal equilibration, and the subsequent 5 ps is used for the generation of the nonadiabatic (NA) Hamiltonian. All NAMD simulations are performed at the $\Gamma$-point using the PBE exchange-correlation functional. The use of a single \textit{k}-point is justified by the direct band gap nature of these antiperovskite nitrides, with both the valence band maximum (VBM) and conduction band minimum (CBM) located at the $\Gamma$-point (see Fig. 1 of SI). To extend the observation window for recombination dynamics, the NA Hamiltonian extracted from this 5 ps segment is iterated multiple times. The first 50 geometries are chosen as initial configurations, and for each geometry, 10,000 stochastic realizations are performed to ensure convergence. SOC effects are excluded from the NAMD simulations, as they do not significantly alter the band gap magnitude (see Table 2 of SI), while introducing a substantial increase in computational cost.

\begin{figure*}
    \centering
    \includegraphics[width=0.85\textwidth]{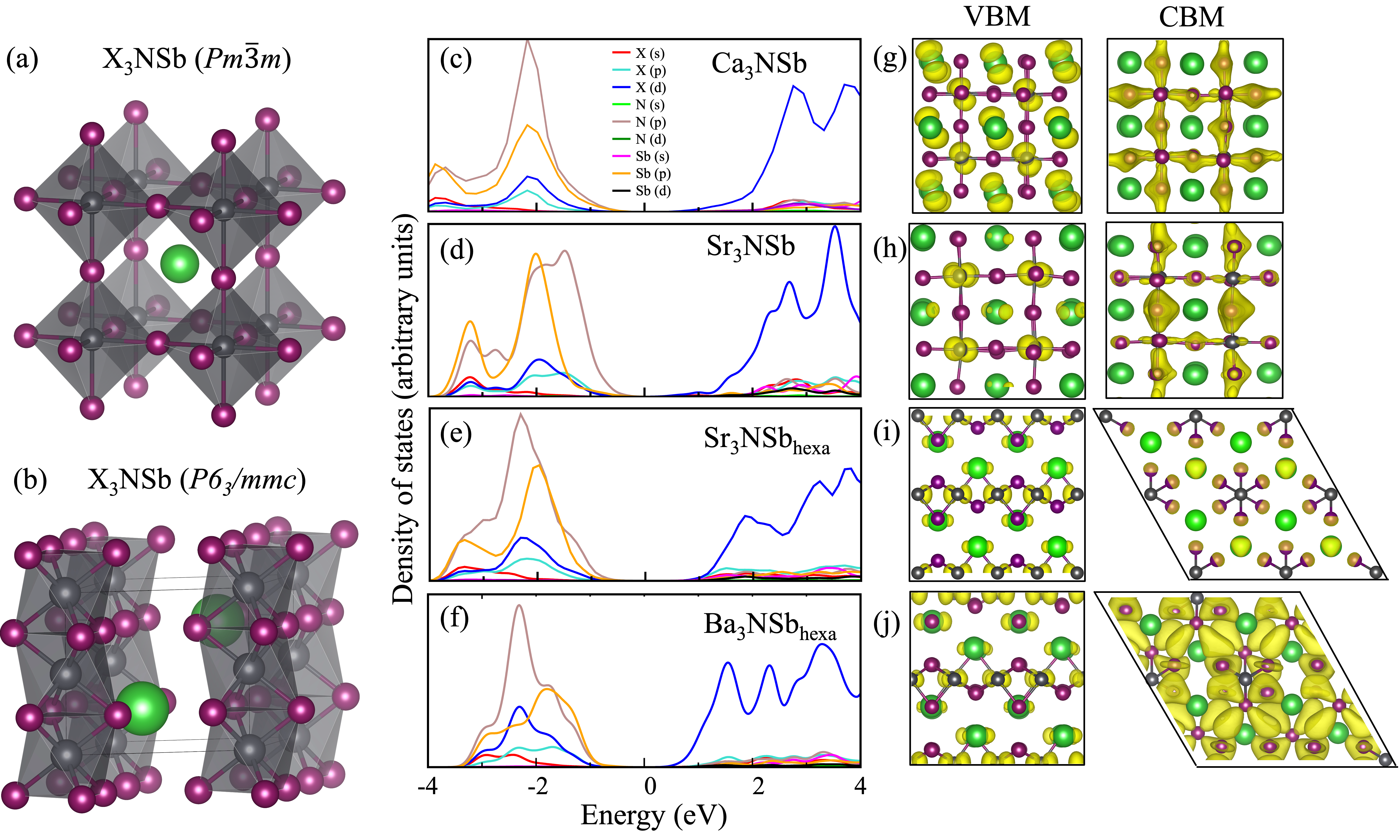}
    \caption{Unit cells of (a) cubic ($\mathit{Pm\overline{3}m}$) $\mathrm{X_3NSb}$ and (b) hexagonal ($\mathit{P6_3/mmc}$) $\mathrm{X_3NSb}$. Panels (c–f) show the atom- and orbital-resolved density of states for the investigated antiperovskites. Panels (g–j) illustrate the corresponding spatial charge distributions of the valence band maximum (VBM) and conduction band minimum (CBM) for $\mathrm{Ca_3NSb}$, $\mathrm{Sr_3NSb}$, $\mathrm{Sr_3NSb_{hexa}}$, and $\mathrm{Ba_3NSb_{hexa}}$, respectively. Grey, purple, and green spheres denote nitrogen, X-site cations, and antimony atoms, respectively.}
    \label{str_dos}
\end{figure*}
The GGA-PBE functional is known to underestimate band gaps. Our previous work \cite{monga2024theoretical} demonstrates that, for Sr$_3$NSb, the quasiparticle $G_0W_0$ approximation on top of HSE06+SOC ($G_0W_0$@HSE06+SOC) yields band gaps in excellent agreement with experiment. However, using hybrid functionals or beyond in NAMD simulations is computationally prohibitive. Therefore, for the present study, we employ the PBE functional, which offers a reasonable balance between computational cost and accuracy, and is widely used for both inorganic and organic perovskites \cite{nayak2025optimizing,wang2025sub,yang2025reducing}. In order to correct for the band gap underestimation in PBE-based NAMD simulations, a scissor operator is applied. The magnitude of the correction corresponds to the difference between the $G_0W_0$@HSE06+SOC and time averaged PBE band gaps, thereby enabling a more realistic estimation of recombination lifetimes. It is important to note that the objective here is to obtain a qualitative understanding of lifetime trends rather than absolute quantitative accuracy. The time-averaged band gaps, obtained by averaging the instantaneous band gaps over a 5 ps MD trajectory, along with the band gaps computed using hybrid HSE06 functional, and the $G_0W_0$ approximation, are reported in Table 2 of SI.

\section{Results}
\subsection{Geometric structure}
Fig.~\ref{str_dos} (a) and (b) show the cubic ($\mathit{Pm\overline{3}m}$) and hexagonal ($\mathit{P6_3/mmc}$) unit cells of X$_3$NSb (X = Ca, Sr, and Ba) antiperovskite nitrides. Ca$_3$NSb and Sr$_3$NSb crystallize in the cubic $\mathit{Pm\overline{3}m}$ phase, whereas $\mathrm{Ba_3NSb}$ adopts a distorted hexagonal $\mathit{P6_3/mmc}$ structure in its ground state (hereafter denoted as $\mathrm{Ba_3NSb_{hexa}}$). In the ideal cubic structure (Fig.\ref{str_dos} (a)), Sb atoms occupy the cuboctahedral center, while N-centered NX$_6$ octahedra are arranged at the corners of the unit cell.

Substituting Ca with Sr at the X-site leads to an increase in the X–N bond length from 2.437 \text{\r{A}} to 2.598 \text{\r{A}}, accompanied by an expansion of the lattice parameters (Table S1 of SI). However, the substantially larger ionic radius of Ba destabilizes the cubic framework through enhanced Ba-N repulsion, favoring a symmetry-lowered hexagonal arrangement. Consequently, $\mathrm{Ba_3NSb_{hexa}}$ exhibits a further increase in the average Ba–N bond length to 2.686~\text{\r{A}}.

Although Sr$_3$NSb adopts the cubic structure in its ground state, we also consider its hexagonal ($\mathit{P6_3/mmc}$) polymorph to enable a direct assessment of symmetry effects at fixed chemical composition. In $\mathrm{Sr_3NSb_{hexa}}$, the average Sr-N bond length slightly decreases to 2.525~\text{\r{A}} relative to the cubic phase, reflecting a modified local coordination environment induced by symmetry lowering. Within the hexagonal phases, X-N bond lengths systematically with increasing X-site cation size.

\subsection{Static electronic properties}\label{elec}
Fig.~\ref{str_dos} (c-j) depicts the atom- and orbital-resolved density of states (DOS) near the band edges together with the real-space charge distributions associated with the VBM and CBM. Across all X$_3$NSb antiperovskites, the VBM is dominated by N and Sb $p$ orbitals, while the CBM primarily arises from the X-site \textit{d} orbitals. In contrast to conventional LHPs, where A-site cations make negligible contributions to the frontier electronic states, the A-site atoms in X$_3$NSb (A=Sb) contribute significantly to the band edges, highlighting the active role of all atomic constituents in these antiperovskites. 

Substituting Ca with Sr in the cubic phase reduces Sb 5$p$$-$N 2$p$ hybridization (Fig.~\ref{str_dos} (d)), likely due to the increased Sb$-$N distances from 4.224 \text{\r{A}} in Ca$_3$NSb to 4.545 \text{\r{A}} in Sr$_3$NSb. In $\mathrm{Sr_3NSb_{hexa}}$, however, reduced symmetry, altered coordination, and slightly shorter Sb–N distances (4.497 \text{\r{A}}) enhance Sb 5$p$–N 2$p$ overlap relative to Sr$_3$NSb. $\mathrm{Ba_3NSb_{hexa}}$ also exhibits strong Sb 5$p$$-$N 2$p$ mixing at the VBM. Remarkably, this enhanced hybridization persists despite the monotonic increase in Sb$-$N distances from Sr$_3$NSb$_{hexa}$ (4.497 \text{\r{A}}) to $\mathrm{Ba_3NSb_{hexa}}$ (4.644 \text{\r{A}}). This indicates that the incorporation of Ba, accompanied by reduced symmetry and local lattice distortions in the hexagonal phase, modifies the bonding environment in a way that strengthens Sb–N hybridization even at longer Sb-N bond distances.
\begin{figure}[h]
    \centering
    \includegraphics[width=0.45\textwidth]{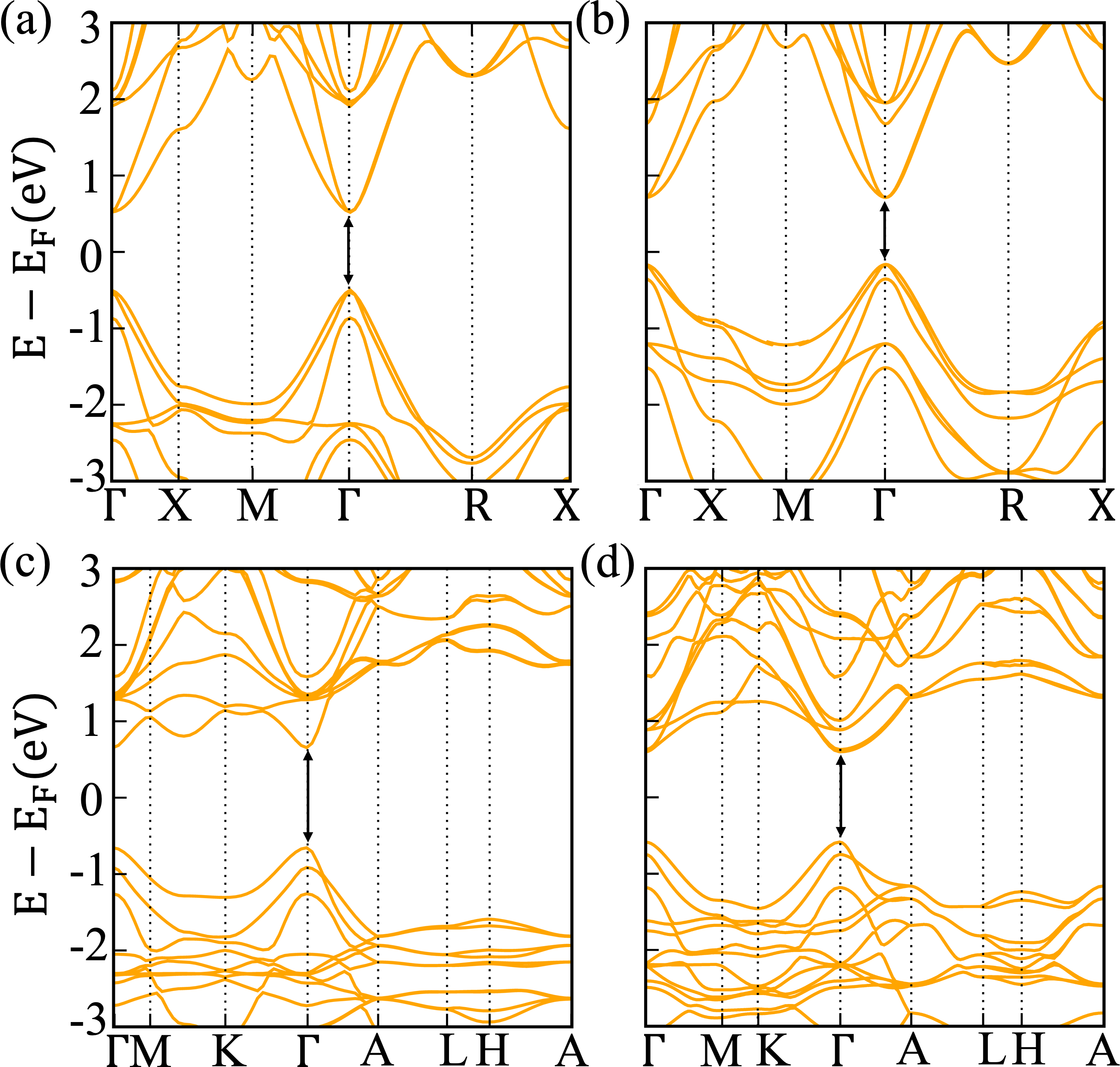}
    \caption{Electronic band structures of (a) $\mathrm{Ca_3NSb}$, (b) $\mathrm{Sr_3NSb}$, (c) $\mathrm{Sr_3NSb_{hexa}}$, and (d) $\mathrm{Ba_3NSb_{hexa}}$ computed using the HSE06 functional with spin–orbit coupling (SOC). All compounds exhibit direct band gaps at the $\Gamma$ point, with values of 1.02 eV, 0.87 eV, 1.31 eV, and 1.18 eV for Ca$_3$NSb, Sr$_3$NSb, $\mathrm{Sr_3NSb_{hexa}}$, and $\mathrm{Ba_3NSb_{hexa}}$, respectively.}
    \label{band_str}
\end{figure}

The electronic band structures calculated using the HSE06 hybrid functional including SOC are presented in Fig.~\ref{band_str}. All compounds exhibit direct band gaps at the $\Gamma$-point, with values of 1.02 eV, 0.87 eV, 1.31 eV and 1.18 eV for Ca$_3$NSb, Sr$_3$NSb, $\mathrm{Sr_3NSb_{hexa}}$, and $\mathrm{Ba_3NSb_{hexa}}$, respectively. Among the cubic phases, the band gap decreases from Ca$_3$NSb to Sr$_3$NSb, consistent with the findings of Mochizuki \textit{et al.}  \cite{mochizuki2020theoretical}, who reported a positive correlation between the band gap and the inverse lattice constant in cubic inorganic antiperovskite nitrides. This trend originates from the lowering of the X $d$ orbital energies with increasing ionic radius, which reduces the CBM energy and narrows the band gap \cite{zhong2021structure}. In $\mathrm{Sr_3NSb_{hexa}}$, the gap increases substantially relative to its cubic counterpart, highlighting the gap-widening effect induced by symmetry-lowering distortions. Similarly, moving from Sr$_3$NSb to $\mathrm{Ba_3NSb_{hexa}}$ results in an increase in the band gap, in contrast to the decreasing trend from Ca$_3$NSb to Sr$_3$NSb with an increase in the X-site cation size. This reversal further highlights that crystal symmetry plays a decisive role in determining the electronic gap. Such distortion-driven band gap widening is well established in conventional perovskites \cite{prasanna2017band, knutson2005tuning, kong2016simultaneous} and is clearly manifested in X$_3$NSb antiperovskites. However, within the hexagonal family, the band gap decreases with increasing X-site cation size, from $\mathrm{Sr_3NSb_{hexa}}$ to $\mathrm{Ba_3NSb_{hexa}}$, mirroring the trend observed in the cubic phase. Thus, both the X-site cation size and the crystal’s structural symmetry play an important role in tuning the electronic properties of X$_3$NSb antiperovskites.

\subsection{Dynamical structural and electronic properties}
AIMD simulations provide valuable insights into the time-dependent structural behavior of these compounds. Upon thermal excitation to 300 K, atomic fluctuations arise that influence electron-phonon interactions. We begin by quantifying the atomic-level fluctuations using the root mean square fluctuation (RMSF) of each species over a 5 ps trajectory (Table ~\ref{rmsf}). Within the cubic phases (Ca$_3$NSb and Sr$_3$NSb), the RMSF of the X-site cations decreases systematically from Ca to Sr, reflecting the suppression of thermal motion with increasing cation mass. In contrast, nitrogen exhibits a monotonic increase in RMSF with increasing X-site ionic radius. This trend is consistent with the progressive elongation of X–N bonds and the accompanying increase in lattice free volume, which permits larger vibrational amplitudes of the lighter N atoms despite the heavier surrounding cations. A comparison between Sr$_3$NSb and its hexagonal polymorph $\mathrm{Sr_3NSb_{hexa}}$ isolates the role of symmetry on atomic dynamics. While the RMSF of the Sr cations remains essentially unchanged across the two phases, the fluctuation of N is slightly increased, whereas that of Sb is reduced in the hexagonal structure. The reduced Sb RMSF indicates that the anisotropic coordination environment in the hexagonal phase imposes additional geometric constraints on the Sb-centered bonding network, thereby limiting its thermal displacement relative to the cubic phase. Notably, the shorter average X-N bond length in $\mathrm{Sr_3NSb_{hexa}}$ does not lead to reduced N fluctuations, indicating that local symmetry breaking, rather than bond length alone, control N dynamics in the hexagonal phase. The same behavior persists in $\mathrm{Ba_3NSb_{hexa}}$, where the heavy Ba cations exhibit the smallest RMSF among all X-site cations, while Sb fluctuations remain comparable to those in Sr$_3$NSb$_{\mathrm{hexa}}$ and the N RMSF is slightly larger. Overall, these results demonstrate that atomic fluctuations in X$_3$NSb antiperovskites arise from an interplay between cation mass, bond geometry, and crystal symmetry.
\begin{table}[h]
    \centering
    \caption{Root mean square fluctuation (RMSF) of the atomic positions for the X-site ion, N, and Sb atoms in \text{\r{A}} for $\mathrm{Ca_3NSb}$, $\mathrm{Sr_3NSb}$, $\mathrm{Sr_3NSb_{hexa}}$, and $\mathrm{Ba_3NSb_{hexa}}$.}
    \label{rmsf}
    \begin{tabular}{l c c c}
    \toprule
        X$_3$NSb & X & N & Sb \\
        \midrule
        Ca$_3$NSb & 0.085 & 0.087 & 0.065 \\
        Sr$_3$NSb & 0.068 & 0.097 & 0.067 \\
        $\mathrm{Sr_3NSb_{hexa}}$ & 0.068 & 0.098 & 0.062 \\
        $\mathrm{Ba_3NSb_{hexa}}$ & 0.057 & 0.099 & 0.062 \\
        \bottomrule
    \end{tabular}
\end{table}

\begin{figure}[h]
    \centering
    \includegraphics[width=0.46\textwidth]{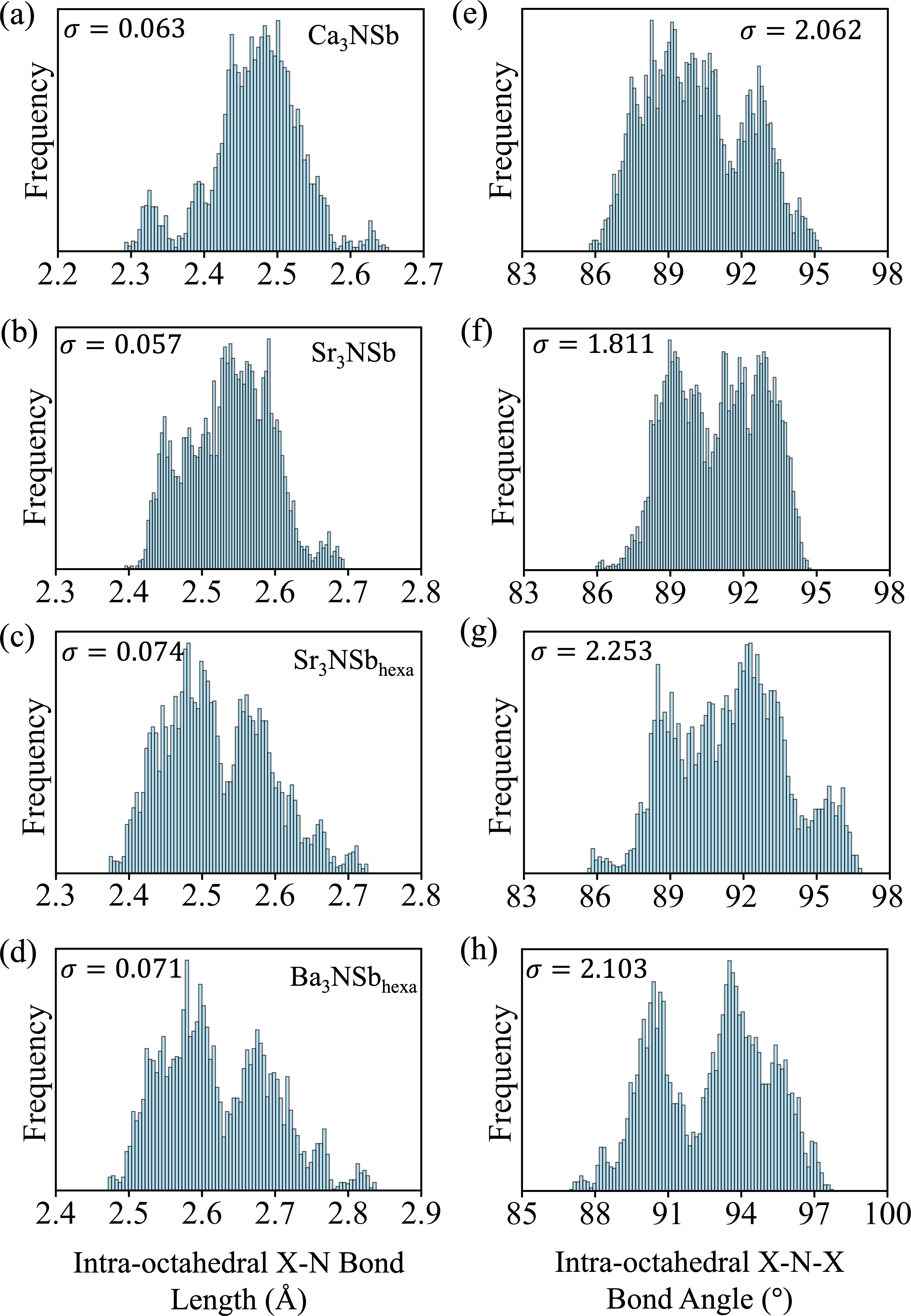}
    \caption{Statistical distribution of (a-d) intra-octahedral X-N bond lengths and (e-h) intra-octahedral X-N-X bond angles, computed using AIMD simulations, for $\mathrm{Ca_3NSb}$ (a, e), $\mathrm{Sr_3NSb}$ (b, f), $\mathrm{Sr_3NSb_{hexa}}$ (c, g) and $\mathrm{Ba_3NSb_{hexa}}$  (d, h). Standard deviations ($\sigma$) of the bond length and bond angle fluctuations are listed.}
    \label{length_angle_fluc}
\end{figure}

Thermal excitation induces a noticeable expansion in the intra-octahedral X-N bond lengths, with time-averaged values of 2.449 \text{\r{A}}, 2.626 \text{\r{A}}, 2.532 \text{\r{A}}, and 2.688 \text{\r{A}} for Ca$_3$NSb, Sr$_3$NSb, $\mathrm{Sr_3NSb_{hexa}}$, and $\mathrm{Ba_3NSb_{hexa}}$, respectively. These values are obtained by averaging all intra-octahedral X-N bond lengths over the entire 5 ps AIMD trajectory. To quantify the variability of these bond lengths, we construct histograms for the X-N distributions, shown in Fig.~\ref{length_angle_fluc} (a-d). The corresponding standard deviations ($\sigma$) reveal a systematic reduction in bond-length fluctuations with increasing X-site cation size within a given symmetry: $\sigma$ decreases from Ca$_3$NSb (0.063 \AA) to Sr$_3$NSb (0.057 \AA), and from $\mathrm{Sr_3NSb_{hexa}}$ (0.074 \AA) to $\mathrm{Ba_3NSb_{hexa}}$ (0.071 \AA). At fixed chemical composition, a clear symmetry effect is observed$-$Sr$_3$NSb exhibits substantially narrower bond-length distributions than its hexagonal polymorph$-$indicating enhanced structural flexibility upon symmetry lowering. A similar behavior is observed for the intra-octahedral X–N–X bond angle distributions (Fig.~\ref{length_angle_fluc}(e–h)). Sr$_3$NSb shows the smallest angular fluctuations ($\sigma$ = 1.811$\degree$), reflecting the rigidity of its NX$_6$ octahedra. While Ca$_3$NSb exhibits larger angular variability ($\sigma$ = 2.062$\degree$), consistent with the lighter Ca$^{2+}$ cation. The hexagonal phases display enhanced angular fluctuations relative to cubic structures, with $\sigma$ = 2.253$\degree$ for $\mathrm{Sr_3NSb_{hexa}}$ and 2.103$\degree$ for $\mathrm{Ba_3NSb_{hexa}}$. Collectively, these results demonstrate that NX$_6$ octahedral dynamics are governed by a combined influence of cation mass and crystallographic symmetry, with symmetry lowering playing a decisive role in enhancing structural fluctuations.
\begin{figure}[h]
    \centering
    \includegraphics[width=0.48\textwidth]{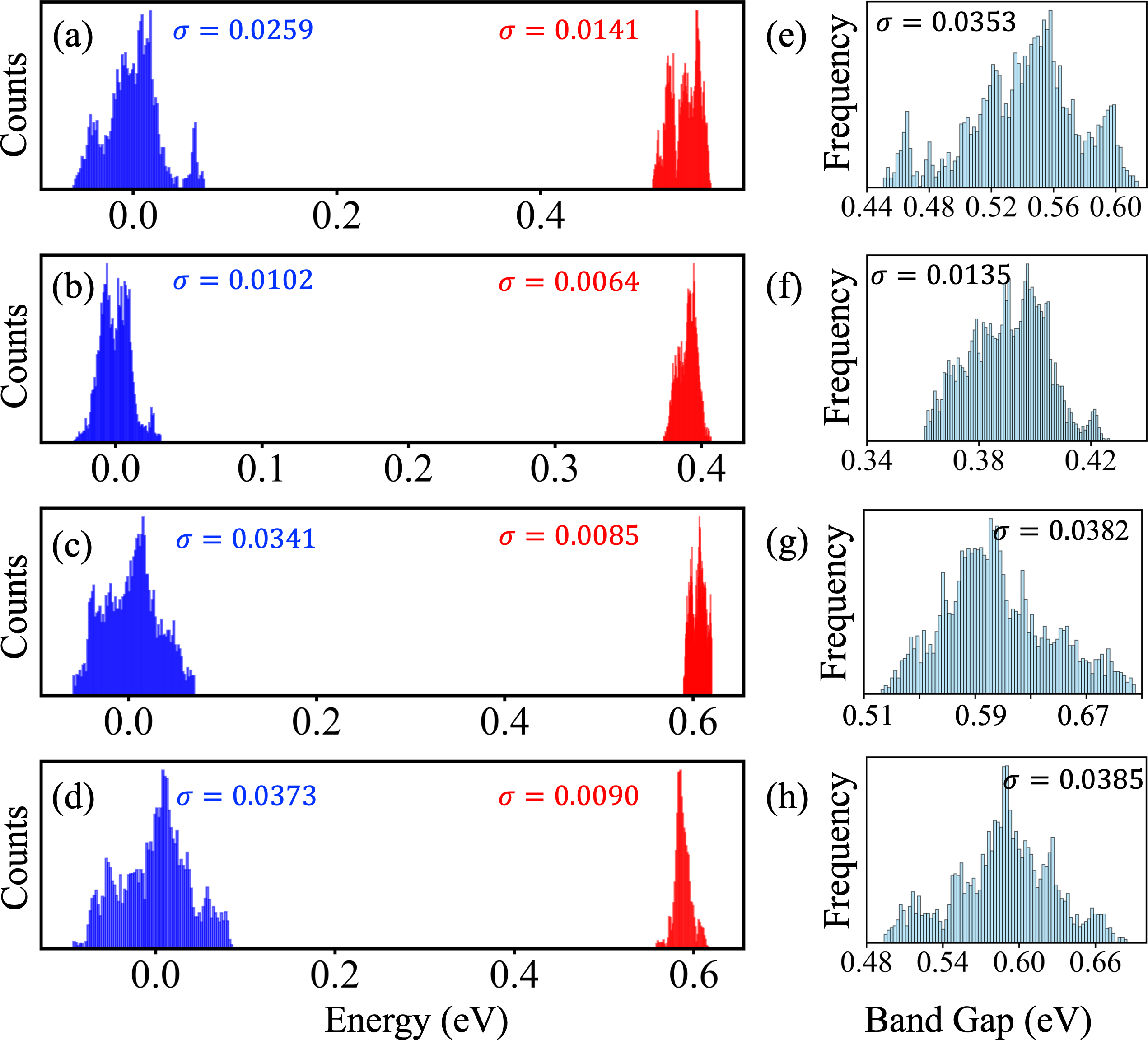}
    \caption{Statistical distribution of (a-d) VBM (blue) and CBM (red), and (e-h) corresponding band gap distributions, computed using AIMD simulations, for $\mathrm{Ca_3NSb}$ (a, e), $\mathrm{Sr_3NSb}$ (b, f), $\mathrm{Sr_3NSb_{hexa}}$ (c, g), and $\mathrm{Ba_3NSb_{hexa}}$  (d, h), respectively. Standard deviations ($\sigma$) of the energy fluctuations are listed.}
    \label{band_edge_fluc} 
\end{figure}

These thermally driven structural fluctuations directly influence the electronic structure through electron–phonon interactions. To quantify this effect, we analyze the instantaneous energies of the VBM and CBM along the AIMD trajectory, with statistical distributions shown in Fig.~\ref{band_edge_fluc}(a–d), together with the corresponding band gap histograms (Fig.~\ref{band_edge_fluc}(e–h)). The data reveal a clear correlation between lattice dynamics and temporal fluctuations of the band-edge eigenvalues. Across all four compositions, the VBM exhibits larger energy fluctuations than the CBM. The standard deviation of the VBM energy follows the trend: $\mathrm{Ba_3NSb_{hexa}}$ (0.0373) $>$ $\mathrm{Sr_3NSb_{hexa}}$ (0.0341) $>$ Ca$_3$NSb (0.0259) $>$ Sr$_3$NSb (0.0102), which mirrors the hierarchy of octahedral flexibility identified earlier. In contrast, the CBM fluctuations follow a distinct order: Ca$_3$NSb (0.0141) $>$ $\mathrm{Ba_3NSb_{hexa}}$ (0.0090) $>$ $\mathrm{Sr_3NSb_{hexa}}$ (0.0085) $>$ Sr$_3$NSb (0.0064), indicating a weaker and qualitatively different sensitivity of the conduction band edge to local structural distortions. 
\begin{table*}
    \centering
    \caption{Canonically averaged scaled band gaps (in eV), with original band gaps included in parenthesis, along with the averaged absolute nonadiabatic (NA) coupling, root mean square (RMS) value of NA coupling, pure-dephasing time, and nonradiative electron-hole recombination lifetime for the VBM-CBM transition in $\mathrm{Ca_3NSb}$, $\mathrm{Sr_3NSb}$, $\mathrm{Sr_3NSb_{hexa}}$, and $\mathrm{Ba_3NSb_{hexa}}$.}
    \label{lifetimes}
    \begin{tabular}{l c c c c c c c c c c}
        \toprule
        X$_3$NSb && Gap && NA coupling && RMS NA && Dephasing && Recombination \\
        && (eV) && (meV) && coupling (meV) && (fs) && (ns) \\
        \midrule
        Ca$_3$NSb && 1.33 (0.54) && 0.57 && 1.02 && 20.3 && 1.36 \\
        Sr$_3$NSb && 1.14 (0.38) && 0.26 && 0.50 && 52.8 && 3.48 \\
        $\mathrm{Sr_3NSb_{hexa}}$ && 1.62 (0.61) && 0.33 && 0.57 && 17.6 && 4.90 \\
        $\mathrm{Ba_3NSb_{hexa}}$ && 1.57 (0.59) && 0.65 && 1.17 && 17.5 && 2.23 \\
        \bottomrule
    \end{tabular}
\end{table*}

Among studied antiperovskites, Sr$_3$NSb exhibits the narrowest energy distributions for both VBM and CBM, resulting in the smallest variation of the instantaneous band gap with a standard deviation of $\sigma = 0.0135$ eV (Fig.~\ref{band_edge_fluc}(e)). In comparison, Ca$_3$NSb, $\mathrm{Sr_3NSb_{hexa}}$ and $\mathrm{Ba_3NSb_{hexa}}$ display significantly larger band gap fluctuations, with standard deviations of approximately 0.0353 eV, 0.0382 eV and 0.0385 eV, respectively. The suppressed band-edge fluctuations in Sr$_3$NSb point to comparatively weaker electron–phonon coupling and reduced nonradiative recombination losses at finite temperature.

\subsection{Electron-vibrational interactions}
The extent of thermal fluctuations in Kohn–Sham orbital energies indicates the strength of electron–vibration interactions. Inelastic electron–phonon scattering, captured by NA coupling, facilitates energy exchange between electrons and vibrations, enabling nonradiative transitions between electronic states. In contrast, elastic scattering disrupts coherence between the initial and final states during an electronic transition. Although it does not involve direct energy exchange, it significantly influences nonradiative recombination, as transitions can only proceed once quantum coherence between the initial and final states is established. The loss of coherence typically acts to suppress or slow down these transitions, as exemplified by the quantum Zeno effect \cite{prezhdo2000quantum,kilina2013quantum}. Thus, both elastic and inelastic processes play an important role in governing  the nonradiative recombination dynamics. 

NA coupling, characterizing inelastic electron-vibration scattering, is calculated as:
\[ d_{ij} = -i\hbar\langle \varphi_i|\nabla_R |\varphi_j \rangle \cdot \dot{R} = -i\hbar \langle \varphi_i | \frac{\partial}{\partial t} | \varphi_j \rangle\]
where, d$_{ij}$ is the NA coupling between the states $i$ and $j$.
$\varphi_i$ and $\varphi_j$ are their corresponding wavefunctions, and $\dot{R}$ denotes the instantaneous nuclear velocity, determined by the kinetic energy, which is proportional to temperature. $\langle \varphi_i|\nabla_R|\varphi_j \rangle$ represents the matrix element which quantifies how electronic wavefunctions depend on atomic displacements and requires the wavefunction of VBM and CBM to be confined within the same spatial area. As shown in Fig.~\ref{str_dos} (g-j), the VBM and CBM states reside primarily on different atomic species, leading to a reduced spatial overlap between electron and hole wavefunctions and, consequently, a suppression of the NA coupling. Nevertheless, the magnitude of NA coupling remains finite due to minor contributions from secondary atomic orbitals. Table ~\ref{lifetimes} summarizes the averaged and root mean squared NA couplings for all compounds. Among them, Sr$_3$NSb exhibits the weakest NA coupling, consistent with its rigid octahedral framework, narrow band-edge distributions, and relatively localized VBM, which altogether the sensitivity of the electronic states to lattice vibrations. Upon transitioning to the hexagonal phase, $\mathrm{Sr_3NSb_{hexa}}$ displays an enhanced NA coupling compared to its cubic counterpart, reflecting the increased structural distortions in the symmetry-lowered bonding environment. However, the coupling remains smaller than in Ca$_3$NSb and $\mathrm{Ba_3NSb_{hexa}}$, which may be attributed to the larger orbital overlap in these structures. Overall, larger NA coupling values signify stronger electron–lattice interactions, which facilitate faster nonradiative recombination processes.

\begin{figure}[h]
    \centering
    \includegraphics[width=0.38\textwidth]{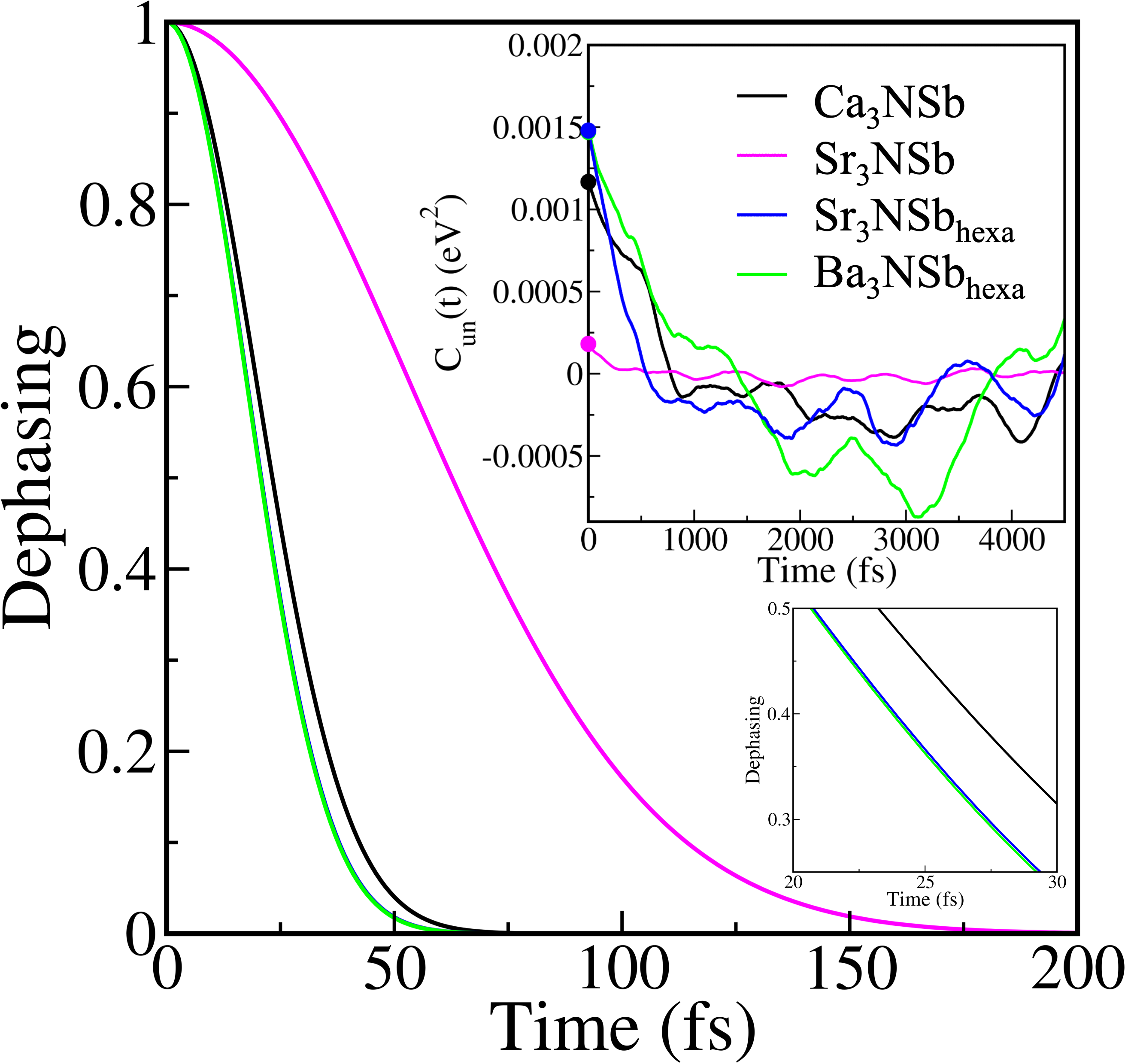}
    \caption{Pure dephasing functions corresponding to the VBM–CBM transition in $\mathrm{Ca_3NSb}$ (black), $\mathrm{Sr_3NSb}$ (magenta), $\mathrm{Sr_3NSb_{hexa}}$ (blue), and $\mathrm{Ba_3NSb_{hexa}}$ (green). The upper inset displays the unnormalized autocorrelation function (un-ACF), whose initial value (indicated by solid circles) represents the square of the band gap fluctuations. A higher initial un-ACF value typically corresponds to a shorter pure dephasing time. The lower inset is the zoomed view of the dephasing function showing almost overlapping curves for $\mathrm{Sr_3NSb_{hexa}}$ (blue), and $\mathrm{Ba_3NSb_{hexa}}$ (green).}
    \label{deph}
\end{figure}
To characterize elastic electron–phonon interactions, we compute the decoherence times, estimated as pure-dephasing time of the optical response theory. Under the second-order cumulant approximation, the pure-dephasing function is expressed as:
\[ D_{ij}(t) = exp\left(\frac{-1}{\hbar^2} \int_0^t dt' \int_0^{t'} dt'' C_{ij}(t'') \right) \]
where $C_{ij}(t) = <\delta E_{ij}(t)\delta E_{ij}(0)>$ is the unnormalized autocorrelation function (un-ACF) of the phonon-induced band gap fluctuation. $E_{ij}(t) = E_{ij}(t) - <E_{ij}>$ represents the fluctuation of the energy gap between states $i$ and $j$, and $<E_{ij}>$ denotes the ensemble-averaged band gap. The pure-dephasing (or decoherence) time is extracted by fitting $D_{ij}(t)$ (Fig.~\ref{deph}) to a Gaussian decay function of the form: $f(t) = exp(-0.5(\frac{t}{\tau})^2)$, with the corresponding values summarized in Table ~\ref{lifetimes}. The initial value of the un-ACF, C$_{ij}$(0), represents the square of band gap fluctuation. The calculated band gap variances (in Fig.~\ref{band_edge_fluc} (e-h)) corroborate the initial values of the un-ACF (inset of Fig.~\ref{deph}), which decrease in the sequence: Sr$_3$NSb $<$ Ca$_3$NSb $<$ $\mathrm{Sr_3NSb_{hexa}}$ $<$ $\mathrm{Ba_3NSb_{hexa}}$. Greater initial value of un-ACF and more asymmetric ACF favor faster dephasing. The pure dephasing times decrease in the opposite order: $\mathrm{Ba_3NSb_{hexa}}$ $<$ $\mathrm{Sr_3NSb_{hexa}}$ $<$ Ca$_3$NSb $<$ Sr$_3$NSb, as expected. The suppression of decoherence/pure dephasing in the Ca$_3$NSb and hexagonal systems ($\mathrm{Sr_3NSb_{hexa}}$ and $\mathrm{Ba_3NSb_{hexa}}$) is attributed to enhanced electronic energy level fluctuations in comparison to Sr$_3$NSb. The pure-dephasing times of X$_3$NSb compounds (17-53 fs) are much shorter than the electron-hole recombination lifetimes, underscoring the importance of including decoherence into the NAMD simulations. 

\subsection{Nonradiative electron-hole recombination}
\begin{figure}[h]
    \centering
    \includegraphics[width=0.38\textwidth]{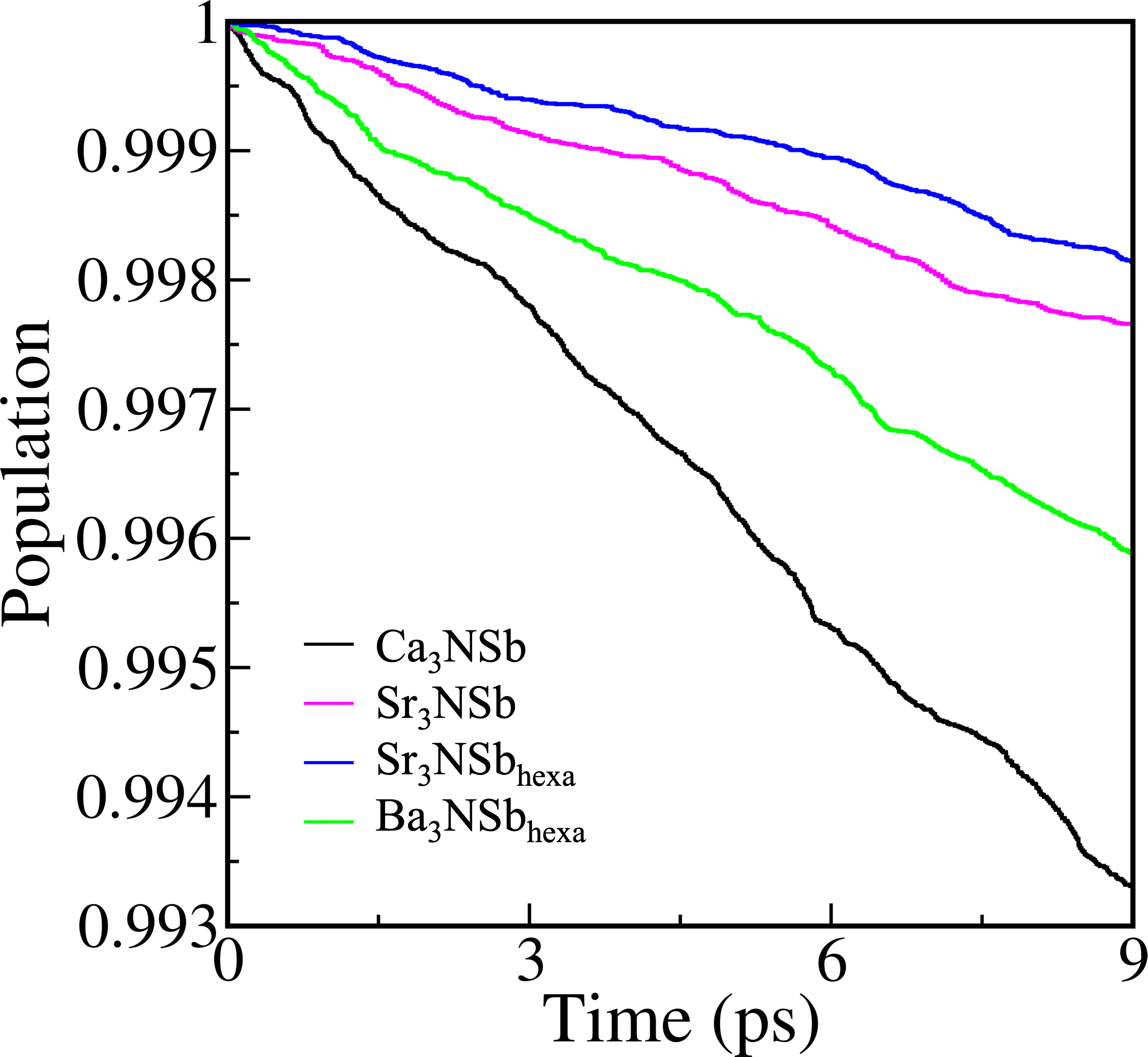}
    \caption{Nonradiative electron–hole recombination dynamics associated with the VBM–CBM transition in $\mathrm{Ca_3NSb}$ (black), $\mathrm{Sr_3NSb}$ (magenta), $\mathrm{Sr_3NSb_{hexa}}$ (blue), and $\mathrm{Ba_3NSb_{hexa}}$ (green).}
    \label{gw_pop.png}
\end{figure}
Fig.~\ref{gw_pop.png} depicts the time evolution of first-excited state populations during nonradiative electron-hole recombination in X$_3$NSb antiperovskites. Since PBE systematically underestimates the band gap, the gaps are rescaled to their corresponding $G_0W_0$@HSE06+SOC values (listed in Table ~\ref{lifetimes}), which reproduce experimental values for Sr$_3$NSb \cite{monga2024theoretical}. Given that intraband relaxation in semiconductors typically occurs on sub-picoseconds timescales, we assume that the electron and hole have already relaxed to the CBM and VBM, respectively. Recombination time constants are extracted by fitting the population decay to an exponential function, $P(t) = \exp\left(-t/\tau\right)$, and the resulting lifetimes are also summarized in Table~\ref{lifetimes}.

NAMD simulations reveal the following trend in recombination lifetimes: $\mathrm{Sr_3NSb_{hexa}}$ (4.90 ns) $>$ Sr$_3$NSb (3.48 ns) $>$ $\mathrm{Ba_3NSb_{hexa}}$ (2.23 ns) $>$ Ca$_3$NSb (1.36 ns). These lifetimes are mainly governed by the combined influence of the band gap, the strength of NA coupling, and  the pure-dephasing time. Larger band gaps and weaker NA coupling, and quick dephasing suppress electronic transitions, thereby reducing nonradiative recombination rates.

Ca$_3$NSb exhibits the shortest nonradiative recombination lifetime, reflecting the combined effect of its small band gap and strong NA coupling. Although $\mathrm{Ba_3NSb_{hexa}}$ shows even stronger NA coupling, its larger band gap and faster electronic dephasing suppress recombination relative to Ca$_3$NSb.
Within the cubic phase, substitution of Ca by Sr localizes the VBM and reduces structural and electronic fluctuations, leading to weaker NA coupling. However, Sr$_3$NSb still exhibits a shorter lifetime than its hexagonal polymorph because its lattice rigidity limit band-gap fluctuations and prolong electronic coherence. $\mathrm{Sr_3NSb_{hexa}}$ displays the longest recombination lifetime, driven by the synergistic combination of a larger band gap, reduced NA coupling, and rapid dephasing. Among the two hexagonal systems, the comparatively stronger NA coupling and smaller band gap of $\mathrm{Ba_3NSb_{hexa}}$ lead to faster nonradiative recombination than in $\mathrm{Sr_3NSb_{hexa}}$.

Overall, these results demonstrate that nonradiative carrier lifetimes in X$_3$NSb antiperovskites are governed by a delicate balance between band gap, lattice-driven NA coupling, and dephasing times. Crucially, symmetry lowering within a fixed chemical composition (Sr$_3$NSb) emerges as an effective strategy for suppressing nonradiative losses, underscoring the importance of structural motifs—beyond chemistry alone—in designing photovoltaic materials.

\section{Conclusion}
Using a combination of NAMD and TDDFT, we investigated the influence of X-site cation substitution and structural symmetry on nonradiative electron-hole recombination dynamics in Ca$_3$NSb,  Sr$_3$NSb, $\mathrm{Sr_3NSb_{hexa}}$, and $\mathrm{Ba_3NSb_{hexa}}$. Band structure calculations reveal that all compounds are direct band gap semiconductors with both the VBM and CBM located at the $\Gamma$ point. In the cubic phase (Ca$_3$NSb and Sr$_3$NSb), increasing X-site cation size narrows the band gap through a downward shift of the X $d$ orbitals that primarily govern the CBM. A similar cation-size dependence of the band gap is observed in the hexagonal phase. However, relative to the cubic structures, lattice distortions in $\mathrm{Sr_3NSb_{hexa}}$ and $\mathrm{Ba_3NSb_{hexa}}$ result in wider band gaps, consistent with trends reported for symmetry-lowered halide perovskites. 

AIMD simulations show that Sr$_3$NSb exhibits the narrowest X-N bond length and X-N-X bond angle distributions, reflecting the rigidity of its NX$_6$ octahedral framework, while Ca$_3$NSb displays moderately enhanced flexibility. The hexagonal phases, $\mathrm{Sr_3NSb_{hexa}}$ and $\mathrm{Ba_3NSb_{hexa}}$, exhibit the largest structural fluctuations due to reduced symmetry. These structural trends are directly reflected in the band gap distribution. The combination of rigid octahedra, narrow band-edge fluctuations, and a relatively localized VBM results in the weakest NA coupling in Sr$_3$NSb. Although NA coupling increases in its hexagonal polymorph, it remains weaker than in Ca$_3$NSb and $\mathrm{Ba_3NSb_{hexa}}$, consistent with its larger band gap and reduced orbital overlap between the VBM and CBM.

Pure-dephasing times, governed by band gap fluctuations, follow the trend Sr$_3$NSb (52.8 fs) $>$ Ca$_3$NSb (20.3 fs) $>$ $\mathrm{Sr_3NSb_{hexa}}$ (17.6 fs) $>$ $\mathrm{Ba_3NSb_{hexa}}$ (17.5 fs), exhibiting an opposite trend to the magnitude of band gap fluctuations. Larger band gap variability accelerates dephasing, as observed in the hexagonal phases relative to the cubic structures. While the two hexagonal systems exhibit comparable dephasing times due to similar band-edge fluctuations, Sr$_3$NSb displays the slowest dephasing overall.

Ultimately, nonradiative recombination lifetimes are governed by the interplay between band gap magnitude, NA coupling strength, and pure-dephasing time. $\mathrm{Sr_3NSb_{hexa}}$ exhibits the longest lifetime due to its wide band gap, weak NA coupling, and rapid dephasing. Its cubic counterpart undergoes faster recombination as a result of its smaller band gap and longer dephasing time. Although $\mathrm{Ba_3NSb_{hexa}}$ and Ca$_3$NSb exhibit comparable NA coupling strengths, the larger band gap and faster dephasing in $\mathrm{Ba_3NSb_{hexa}}$ lead to slower recombination relative to Ca$_3$NSb. Overall, both X-site cation chemistry and structural symmetry play decisive roles in shaping structural dynamics, electronic fluctuations, and nonradiative recombination in X$_3$NSb antiperovskites. These findings establish clear design principles for achieving long-lived excited states through symmetry and lattice engineering.

\section{Acknowledgment}
S.M. acknowledge IIT Delhi for the senior research fellowship. S.B. acknowledge financial support from SERB under a core research grant [Grant no. CRG/2019/000647] to set up his High Performance Computing (HPC) facility ``Veena'' at IIT Delhi for computational resources.

\section{Data Availability}
Data available from the authors upon reasonable request.


\begin{thebibliography}{47}%
	\makeatletter
	\providecommand \@ifxundefined [1]{%
		\@ifx{#1\undefined}
	}%
	\providecommand \@ifnum [1]{%
		\ifnum #1\expandafter \@firstoftwo
		\else \expandafter \@secondoftwo
		\fi
	}%
	\providecommand \@ifx [1]{%
		\ifx #1\expandafter \@firstoftwo
		\else \expandafter \@secondoftwo
		\fi
	}%
	\providecommand \natexlab [1]{#1}%
	\providecommand \enquote  [1]{``#1''}%
	\providecommand \bibnamefont  [1]{#1}%
	\providecommand \bibfnamefont [1]{#1}%
	\providecommand \citenamefont [1]{#1}%
	\providecommand \href@noop [0]{\@secondoftwo}%
	\providecommand \href [0]{\begingroup \@sanitize@url \@href}%
	\providecommand \@href[1]{\@@startlink{#1}\@@href}%
	\providecommand \@@href[1]{\endgroup#1\@@endlink}%
	\providecommand \@sanitize@url [0]{\catcode `\\12\catcode `\$12\catcode
		`\&12\catcode `\#12\catcode `\^12\catcode `\_12\catcode `\%12\relax}%
	\providecommand \@@startlink[1]{}%
	\providecommand \@@endlink[0]{}%
	\providecommand \url  [0]{\begingroup\@sanitize@url \@url }%
	\providecommand \@url [1]{\endgroup\@href {#1}{\urlprefix }}%
	\providecommand \urlprefix  [0]{URL }%
	\providecommand \Eprint [0]{\href }%
	\providecommand \doibase [0]{https://doi.org/}%
	\providecommand \selectlanguage [0]{\@gobble}%
	\providecommand \bibinfo  [0]{\@secondoftwo}%
	\providecommand \bibfield  [0]{\@secondoftwo}%
	\providecommand \translation [1]{[#1]}%
	\providecommand \BibitemOpen [0]{}%
	\providecommand \bibitemStop [0]{}%
	\providecommand \bibitemNoStop [0]{.\EOS\space}%
	\providecommand \EOS [0]{\spacefactor3000\relax}%
	\providecommand \BibitemShut  [1]{\csname bibitem#1\endcsname}%
	\let\auto@bib@innerbib\@empty
	\bibitem [{\citenamefont {Green}\ \emph {et~al.}(2014)\citenamefont {Green},
		\citenamefont {Ho-Baillie},\ and\ \citenamefont {Snaith}}]{Green2014}%
	\BibitemOpen
	\bibfield  {author} {\bibinfo {author} {\bibfnamefont {M.~A.}\ \bibnamefont
			{Green}}, \bibinfo {author} {\bibfnamefont {A.}~\bibnamefont {Ho-Baillie}},\
		and\ \bibinfo {author} {\bibfnamefont {H.~J.}\ \bibnamefont {Snaith}},\
	}\bibfield  {title} {\bibinfo {title} {The emergence of perovskite solar
			cells},\ }\href@noop {} {\bibfield  {journal} {\bibinfo  {journal} {Nat.
				Photonics}\ }\textbf {\bibinfo {volume} {8}},\ \bibinfo {pages} {506}
		(\bibinfo {year} {2014})}\BibitemShut {NoStop}%
	\bibitem [{\citenamefont {Yin}\ \emph {et~al.}(2015)\citenamefont {Yin},
		\citenamefont {Yang}, \citenamefont {Kang}, \citenamefont {Yan},\ and\
		\citenamefont {Wei}}]{Yin2015}%
	\BibitemOpen
	\bibfield  {author} {\bibinfo {author} {\bibfnamefont {W.-J.}\ \bibnamefont
			{Yin}}, \bibinfo {author} {\bibfnamefont {J.-H.}\ \bibnamefont {Yang}},
		\bibinfo {author} {\bibfnamefont {J.}~\bibnamefont {Kang}}, \bibinfo {author}
		{\bibfnamefont {Y.}~\bibnamefont {Yan}},\ and\ \bibinfo {author}
		{\bibfnamefont {S.-H.}\ \bibnamefont {Wei}},\ }\bibfield  {title} {\bibinfo
		{title} {Halide perovskite materials for solar cells: a theoretical review},\
	}\href@noop {} {\bibfield  {journal} {\bibinfo  {journal} {J. Mater. Chem.
				A}\ }\textbf {\bibinfo {volume} {3}},\ \bibinfo {pages} {8926} (\bibinfo
		{year} {2015})}\BibitemShut {NoStop}%
	\bibitem [{\citenamefont {Kim}\ \emph {et~al.}(2020)\citenamefont {Kim},
		\citenamefont {Lee}, \citenamefont {Jung}, \citenamefont {Shin},\ and\
		\citenamefont {Park}}]{kim2020high}%
	\BibitemOpen
	\bibfield  {author} {\bibinfo {author} {\bibfnamefont {J.~Y.}\ \bibnamefont
			{Kim}}, \bibinfo {author} {\bibfnamefont {J.-W.}\ \bibnamefont {Lee}},
		\bibinfo {author} {\bibfnamefont {H.~S.}\ \bibnamefont {Jung}}, \bibinfo
		{author} {\bibfnamefont {H.}~\bibnamefont {Shin}},\ and\ \bibinfo {author}
		{\bibfnamefont {N.-G.}\ \bibnamefont {Park}},\ }\bibfield  {title} {\bibinfo
		{title} {High-efficiency perovskite solar cells},\ }\href@noop {} {\bibfield
		{journal} {\bibinfo  {journal} {Chem. Rev.}\ }\textbf {\bibinfo {volume}
			{120}},\ \bibinfo {pages} {7867} (\bibinfo {year} {2020})}\BibitemShut
	{NoStop}%
	\bibitem [{\citenamefont {Yu}\ \emph {et~al.}(2020)\citenamefont {Yu},
		\citenamefont {Tsao}, \citenamefont {Zhang},\ and\ \citenamefont
		{Gao}}]{yu2020miscellaneous}%
	\BibitemOpen
	\bibfield  {author} {\bibinfo {author} {\bibfnamefont {X.}~\bibnamefont
			{Yu}}, \bibinfo {author} {\bibfnamefont {H.~N.}\ \bibnamefont {Tsao}},
		\bibinfo {author} {\bibfnamefont {Z.}~\bibnamefont {Zhang}},\ and\ \bibinfo
		{author} {\bibfnamefont {P.}~\bibnamefont {Gao}},\ }\bibfield  {title}
	{\bibinfo {title} {Miscellaneous and perspicacious: Hybrid halide perovskite
			materials based photodetectors and sensors},\ }\href@noop {} {\bibfield
		{journal} {\bibinfo  {journal} {Adv. Opt. Mater.}\ }\textbf {\bibinfo
			{volume} {8}},\ \bibinfo {pages} {2001095} (\bibinfo {year}
		{2020})}\BibitemShut {NoStop}%
	\bibitem [{nre()}]{nrel}%
	\BibitemOpen
	\href@noop {} {\bibinfo {title} {National renewable energy laboratory (nrel).
			best research-cell efficiency chart.}},\ \bibinfo {howpublished}
	{\url{https://www.nrel.gov/pv/cell-efficiency.html}},\ \bibinfo {note}
	{(accessed: January 26, 2021)}\BibitemShut {NoStop}%
	\bibitem [{\citenamefont {Schileo}\ and\ \citenamefont
		{Grancini}(2021)}]{schileo2021lead}%
	\BibitemOpen
	\bibfield  {author} {\bibinfo {author} {\bibfnamefont {G.}~\bibnamefont
			{Schileo}}\ and\ \bibinfo {author} {\bibfnamefont {G.}~\bibnamefont
			{Grancini}},\ }\bibfield  {title} {\bibinfo {title} {Lead or no lead?
			availability, toxicity, sustainability and environmental impact of lead-free
			perovskite solar cells},\ }\href@noop {} {\bibfield  {journal} {\bibinfo
			{journal} {J. Mater. Chem. C}\ }\textbf {\bibinfo {volume} {9}},\ \bibinfo
		{pages} {67} (\bibinfo {year} {2021})}\BibitemShut {NoStop}%
	\bibitem [{\citenamefont {Ju}\ \emph {et~al.}(2018)\citenamefont {Ju},
		\citenamefont {Chen}, \citenamefont {Zhou}, \citenamefont {Dai},
		\citenamefont {Ma}, \citenamefont {Padture},\ and\ \citenamefont
		{Zeng}}]{ju2018toward}%
	\BibitemOpen
	\bibfield  {author} {\bibinfo {author} {\bibfnamefont {M.-G.}\ \bibnamefont
			{Ju}}, \bibinfo {author} {\bibfnamefont {M.}~\bibnamefont {Chen}}, \bibinfo
		{author} {\bibfnamefont {Y.}~\bibnamefont {Zhou}}, \bibinfo {author}
		{\bibfnamefont {J.}~\bibnamefont {Dai}}, \bibinfo {author} {\bibfnamefont
			{L.}~\bibnamefont {Ma}}, \bibinfo {author} {\bibfnamefont {N.~P.}\
			\bibnamefont {Padture}},\ and\ \bibinfo {author} {\bibfnamefont {X.~C.}\
			\bibnamefont {Zeng}},\ }\bibfield  {title} {\bibinfo {title} {Toward
			eco-friendly and stable perovskite materials for photovoltaics},\ }\href@noop
	{} {\bibfield  {journal} {\bibinfo  {journal} {Joule}\ }\textbf {\bibinfo
			{volume} {2}},\ \bibinfo {pages} {1231} (\bibinfo {year} {2018})}\BibitemShut
	{NoStop}%
	\bibitem [{\citenamefont {Chi}\ \emph {et~al.}(2002)\citenamefont {Chi},
		\citenamefont {Kim}, \citenamefont {Hur},\ and\ \citenamefont
		{Jung}}]{chi2002new}%
	\BibitemOpen
	\bibfield  {author} {\bibinfo {author} {\bibfnamefont {E.}~\bibnamefont
			{Chi}}, \bibinfo {author} {\bibfnamefont {W.}~\bibnamefont {Kim}}, \bibinfo
		{author} {\bibfnamefont {N.}~\bibnamefont {Hur}},\ and\ \bibinfo {author}
		{\bibfnamefont {D.}~\bibnamefont {Jung}},\ }\bibfield  {title} {\bibinfo
		{title} {New mg-based antiperovskites $\mathrm{PnNMg_3}$ ($\mathrm{Pn= As,
				Sb}$)},\ }\href@noop {} {\bibfield  {journal} {\bibinfo  {journal} {Solid
				State Commun.}\ }\textbf {\bibinfo {volume} {121}},\ \bibinfo {pages} {309}
		(\bibinfo {year} {2002})}\BibitemShut {NoStop}%
	\bibitem [{\citenamefont {Beznosikov}(2003)}]{beznosikov2003predicted}%
	\BibitemOpen
	\bibfield  {author} {\bibinfo {author} {\bibfnamefont {B.}~\bibnamefont
			{Beznosikov}},\ }\bibfield  {title} {\bibinfo {title} {Predicted nitrides
			with an antiperovskite structure},\ }\href@noop {} {\bibfield  {journal}
		{\bibinfo  {journal} {J. Struct. Chem.}\ }\textbf {\bibinfo {volume} {44}},\
		\bibinfo {pages} {885} (\bibinfo {year} {2003})}\BibitemShut {NoStop}%
	\bibitem [{\citenamefont {Heinselman}\ \emph {et~al.}(2019)\citenamefont
		{Heinselman}, \citenamefont {Lany}, \citenamefont {Perkins}, \citenamefont
		{Talley},\ and\ \citenamefont {Zakutayev}}]{heinselman2019thin}%
	\BibitemOpen
	\bibfield  {author} {\bibinfo {author} {\bibfnamefont {K.~N.}\ \bibnamefont
			{Heinselman}}, \bibinfo {author} {\bibfnamefont {S.}~\bibnamefont {Lany}},
		\bibinfo {author} {\bibfnamefont {J.~D.}\ \bibnamefont {Perkins}}, \bibinfo
		{author} {\bibfnamefont {K.~R.}\ \bibnamefont {Talley}},\ and\ \bibinfo
		{author} {\bibfnamefont {A.}~\bibnamefont {Zakutayev}},\ }\bibfield  {title}
	{\bibinfo {title} {Thin film synthesis of semiconductors in the
			$\mathrm{Mg-Sb-N}$ materials system},\ }\href@noop {} {\bibfield  {journal}
		{\bibinfo  {journal} {Chem. Mater.}\ }\textbf {\bibinfo {volume} {31}},\
		\bibinfo {pages} {8717} (\bibinfo {year} {2019})}\BibitemShut {NoStop}%
	\bibitem [{\citenamefont {Mochizuki}\ \emph {et~al.}(2020)\citenamefont
		{Mochizuki}, \citenamefont {Sung}, \citenamefont {Takahashi}, \citenamefont
		{Kumagai},\ and\ \citenamefont {Oba}}]{mochizuki2020theoretical}%
	\BibitemOpen
	\bibfield  {author} {\bibinfo {author} {\bibfnamefont {Y.}~\bibnamefont
			{Mochizuki}}, \bibinfo {author} {\bibfnamefont {H.-J.}\ \bibnamefont {Sung}},
		\bibinfo {author} {\bibfnamefont {A.}~\bibnamefont {Takahashi}}, \bibinfo
		{author} {\bibfnamefont {Y.}~\bibnamefont {Kumagai}},\ and\ \bibinfo {author}
		{\bibfnamefont {F.}~\bibnamefont {Oba}},\ }\bibfield  {title} {\bibinfo
		{title} {Theoretical exploration of mixed-anion antiperovskite semiconductors
			$\mathrm{M_3XN}$ ($\mathrm{M= Mg, Ca, Sr, Ba; X= P, As, Sb, Bi}$)},\
	}\href@noop {} {\bibfield  {journal} {\bibinfo  {journal} {Phys. Rev.
				Mater.}\ }\textbf {\bibinfo {volume} {4}},\ \bibinfo {pages} {044601}
		(\bibinfo {year} {2020})}\BibitemShut {NoStop}%
	\bibitem [{\citenamefont {Zhong}\ \emph {et~al.}(2021)\citenamefont {Zhong},
		\citenamefont {Feng}, \citenamefont {Wang}, \citenamefont {Han},
		\citenamefont {Yu}, \citenamefont {Xiong}, \citenamefont {Li}, \citenamefont
		{Yang}, \citenamefont {Tang},\ and\ \citenamefont
		{Yuan}}]{zhong2021structure}%
	\BibitemOpen
	\bibfield  {author} {\bibinfo {author} {\bibfnamefont {H.}~\bibnamefont
			{Zhong}}, \bibinfo {author} {\bibfnamefont {C.}~\bibnamefont {Feng}},
		\bibinfo {author} {\bibfnamefont {H.}~\bibnamefont {Wang}}, \bibinfo {author}
		{\bibfnamefont {D.}~\bibnamefont {Han}}, \bibinfo {author} {\bibfnamefont
			{G.}~\bibnamefont {Yu}}, \bibinfo {author} {\bibfnamefont {W.}~\bibnamefont
			{Xiong}}, \bibinfo {author} {\bibfnamefont {Y.}~\bibnamefont {Li}}, \bibinfo
		{author} {\bibfnamefont {M.}~\bibnamefont {Yang}}, \bibinfo {author}
		{\bibfnamefont {G.}~\bibnamefont {Tang}},\ and\ \bibinfo {author}
		{\bibfnamefont {S.}~\bibnamefont {Yuan}},\ }\bibfield  {title} {\bibinfo
		{title} {Structure--composition--property relationships in antiperovskite
			nitrides: Guiding a rational alloy design},\ }\href@noop {} {\bibfield
		{journal} {\bibinfo  {journal} {ACS Appl. Mater. Interfaces}\ }\textbf
		{\bibinfo {volume} {13}},\ \bibinfo {pages} {48516} (\bibinfo {year}
		{2021})}\BibitemShut {NoStop}%
	\bibitem [{\citenamefont {Wang}\ \emph {et~al.}(2020)\citenamefont {Wang},
		\citenamefont {Zhang}, \citenamefont {Zhu}, \citenamefont {L{\"u}},
		\citenamefont {Li}, \citenamefont {Zou},\ and\ \citenamefont
		{Zhao}}]{wang2020antiperovskites}%
	\BibitemOpen
	\bibfield  {author} {\bibinfo {author} {\bibfnamefont {Y.}~\bibnamefont
			{Wang}}, \bibinfo {author} {\bibfnamefont {H.}~\bibnamefont {Zhang}},
		\bibinfo {author} {\bibfnamefont {J.}~\bibnamefont {Zhu}}, \bibinfo {author}
		{\bibfnamefont {X.}~\bibnamefont {L{\"u}}}, \bibinfo {author} {\bibfnamefont
			{S.}~\bibnamefont {Li}}, \bibinfo {author} {\bibfnamefont {R.}~\bibnamefont
			{Zou}},\ and\ \bibinfo {author} {\bibfnamefont {Y.}~\bibnamefont {Zhao}},\
	}\bibfield  {title} {\bibinfo {title} {Antiperovskites with exceptional
			functionalities},\ }\href@noop {} {\bibfield  {journal} {\bibinfo  {journal}
			{Adv. Mater.}\ }\textbf {\bibinfo {volume} {32}},\ \bibinfo {pages} {1905007}
		(\bibinfo {year} {2020})}\BibitemShut {NoStop}%
	\bibitem [{\citenamefont {Dai}\ \emph {et~al.}(2019)\citenamefont {Dai},
		\citenamefont {Ju}, \citenamefont {Ma},\ and\ \citenamefont
		{Zeng}}]{dai2019bi}%
	\BibitemOpen
	\bibfield  {author} {\bibinfo {author} {\bibfnamefont {J.}~\bibnamefont
			{Dai}}, \bibinfo {author} {\bibfnamefont {M.-G.}\ \bibnamefont {Ju}},
		\bibinfo {author} {\bibfnamefont {L.}~\bibnamefont {Ma}},\ and\ \bibinfo
		{author} {\bibfnamefont {X.~C.}\ \bibnamefont {Zeng}},\ }\bibfield  {title}
	{\bibinfo {title} {$\mathrm{Bi (Sb) NCa_3}$: Expansion of perovskite
			photovoltaics into all-inorganic anti-perovskite materials},\ }\href@noop {}
	{\bibfield  {journal} {\bibinfo  {journal} {J. Phys. Chem. C}\ }\textbf
		{\bibinfo {volume} {123}},\ \bibinfo {pages} {6363} (\bibinfo {year}
		{2019})}\BibitemShut {NoStop}%
	\bibitem [{\citenamefont {Kang}(2022)}]{kang2022antiperovskite}%
	\BibitemOpen
	\bibfield  {author} {\bibinfo {author} {\bibfnamefont {Y.}~\bibnamefont
			{Kang}},\ }\bibfield  {title} {\bibinfo {title} {Antiperovskite
			$\mathrm{Sr_3MN}$ and $\mathrm{Ba_3MN}$ ($\mathrm{M= Sb}$ or $\mathrm{Bi}$)
			as promising photovoltaic absorbers for thin-film solar cells: A
			first-principles study},\ }\href@noop {} {\bibfield  {journal} {\bibinfo
			{journal} {J. Am. Ceram. Soc.}\ }\textbf {\bibinfo {volume} {105}},\ \bibinfo
		{pages} {5807} (\bibinfo {year} {2022})}\BibitemShut {NoStop}%
	\bibitem [{\citenamefont {G{\"a}bler}\ \emph {et~al.}(2004)\citenamefont
		{G{\"a}bler}, \citenamefont {Kirchner}, \citenamefont {Schnelle},
		\citenamefont {Schwarz}, \citenamefont {Schmitt}, \citenamefont {Rosner},\
		and\ \citenamefont {Niewa}}]{gabler2004sr3n}%
	\BibitemOpen
	\bibfield  {author} {\bibinfo {author} {\bibfnamefont {F.}~\bibnamefont
			{G{\"a}bler}}, \bibinfo {author} {\bibfnamefont {M.}~\bibnamefont
			{Kirchner}}, \bibinfo {author} {\bibfnamefont {W.}~\bibnamefont {Schnelle}},
		\bibinfo {author} {\bibfnamefont {U.}~\bibnamefont {Schwarz}}, \bibinfo
		{author} {\bibfnamefont {M.}~\bibnamefont {Schmitt}}, \bibinfo {author}
		{\bibfnamefont {H.}~\bibnamefont {Rosner}},\ and\ \bibinfo {author}
		{\bibfnamefont {R.}~\bibnamefont {Niewa}},\ }\bibfield  {title} {\bibinfo
		{title} {{(Sr$_3$N)E and (Ba$_3$N)E (E = Sb, Bi): Synthesis, crystal
				structures, and physical properties}},\ }\href@noop {} {\bibfield  {journal}
		{\bibinfo  {journal} {Z Anorg Allg Chem}\ }\textbf {\bibinfo {volume}
			{630}},\ \bibinfo {pages} {2292} (\bibinfo {year} {2004})}\BibitemShut
	{NoStop}%
	\bibitem [{\citenamefont {Monga}\ \emph {et~al.}(2024)\citenamefont {Monga},
		\citenamefont {Jain}, \citenamefont {Draxl},\ and\ \citenamefont
		{Bhattacharya}}]{monga2024theoretical}%
	\BibitemOpen
	\bibfield  {author} {\bibinfo {author} {\bibfnamefont {S.}~\bibnamefont
			{Monga}}, \bibinfo {author} {\bibfnamefont {M.}~\bibnamefont {Jain}},
		\bibinfo {author} {\bibfnamefont {C.}~\bibnamefont {Draxl}},\ and\ \bibinfo
		{author} {\bibfnamefont {S.}~\bibnamefont {Bhattacharya}},\ }\bibfield
	{title} {\bibinfo {title} {{Theoretical insights into inorganic
				antiperovskite nitrides (X$_3$NA: X= Mg, Ca, Sr, Ba; A= As, Sb): An emerging
				class of materials for photovoltaics}},\ }\href@noop {} {\bibfield  {journal}
		{\bibinfo  {journal} {Phys. Rev. Mater.}\ }\textbf {\bibinfo {volume} {8}},\
		\bibinfo {pages} {105403} (\bibinfo {year} {2024})}\BibitemShut {NoStop}%
	\bibitem [{\citenamefont {Frost}(2017)}]{Frost2017}%
	\BibitemOpen
	\bibfield  {author} {\bibinfo {author} {\bibfnamefont {J.~M.}\ \bibnamefont
			{Frost}},\ }\bibfield  {title} {\bibinfo {title} {Calculating polaron
			mobility in halide perovskites},\ }\href@noop {} {\bibfield  {journal}
		{\bibinfo  {journal} {Phys. Rev. B.}\ }\textbf {\bibinfo {volume} {96}}
		(\bibinfo {year} {2017})}\BibitemShut {NoStop}%
	\bibitem [{\citenamefont {Wang}\ \emph
		{et~al.}(2025{\natexlab{a}})\citenamefont {Wang}, \citenamefont {Dong},
		\citenamefont {Yuan}, \citenamefont {Wen}, \citenamefont {He}, \citenamefont
		{Tong},\ and\ \citenamefont {Prezhdo}}]{wang2025self}%
	\BibitemOpen
	\bibfield  {author} {\bibinfo {author} {\bibfnamefont {K.-P.}\ \bibnamefont
			{Wang}}, \bibinfo {author} {\bibfnamefont {X.}~\bibnamefont {Dong}}, \bibinfo
		{author} {\bibfnamefont {J.-Z.}\ \bibnamefont {Yuan}}, \bibinfo {author}
		{\bibfnamefont {B.}~\bibnamefont {Wen}}, \bibinfo {author} {\bibfnamefont
			{J.}~\bibnamefont {He}}, \bibinfo {author} {\bibfnamefont {C.-J.}\
			\bibnamefont {Tong}},\ and\ \bibinfo {author} {\bibfnamefont {O.~V.}\
			\bibnamefont {Prezhdo}},\ }\bibfield  {title} {\bibinfo {title}
		{{Self-Passivation at the SnO$_2$/Perovskite Interface}},\ }\href@noop {}
	{\bibfield  {journal} {\bibinfo  {journal} {ACS Energy Lett.}\ }\textbf
		{\bibinfo {volume} {10}},\ \bibinfo {pages} {1466} (\bibinfo {year}
		{2025}{\natexlab{a}})}\BibitemShut {NoStop}%
	\bibitem [{\citenamefont {Wang}\ \emph {et~al.}(2024)\citenamefont {Wang},
		\citenamefont {Wu}, \citenamefont {Wang}, \citenamefont {Xu}, \citenamefont
		{He}, \citenamefont {Wen}, \citenamefont {Tong}, \citenamefont {Liu},\ and\
		\citenamefont {Prezhdo}}]{wang2024detrimental}%
	\BibitemOpen
	\bibfield  {author} {\bibinfo {author} {\bibfnamefont {K.-P.}\ \bibnamefont
			{Wang}}, \bibinfo {author} {\bibfnamefont {Z.-W.}\ \bibnamefont {Wu}},
		\bibinfo {author} {\bibfnamefont {K.-F.}\ \bibnamefont {Wang}}, \bibinfo
		{author} {\bibfnamefont {H.-T.}\ \bibnamefont {Xu}}, \bibinfo {author}
		{\bibfnamefont {J.}~\bibnamefont {He}}, \bibinfo {author} {\bibfnamefont
			{B.}~\bibnamefont {Wen}}, \bibinfo {author} {\bibfnamefont {C.-J.}\
			\bibnamefont {Tong}}, \bibinfo {author} {\bibfnamefont {L.-M.}\ \bibnamefont
			{Liu}},\ and\ \bibinfo {author} {\bibfnamefont {O.~V.}\ \bibnamefont
			{Prezhdo}},\ }\bibfield  {title} {\bibinfo {title} {{Detrimental Defect
				Cooperativity at TiO$_2$/CH$_3$NH$_3$PbI$_3$ Interface: Decreased Stability,
				Enhanced Ion Diffusion, and Reduced Charge Lifetime and Transport}},\
	}\href@noop {} {\bibfield  {journal} {\bibinfo  {journal} {ACS Energy Lett.}\
		}\textbf {\bibinfo {volume} {9}},\ \bibinfo {pages} {5888} (\bibinfo {year}
		{2024})}\BibitemShut {NoStop}%
	\bibitem [{\citenamefont {Panigrahi}\ \emph {et~al.}(2022)\citenamefont
		{Panigrahi}, \citenamefont {Sk}, \citenamefont {Jana}, \citenamefont {Ghosh},
		\citenamefont {Deuermeier}, \citenamefont {Martins},\ and\ \citenamefont
		{Fortunato}}]{panigrahi2022tailoring}%
	\BibitemOpen
	\bibfield  {author} {\bibinfo {author} {\bibfnamefont {S.}~\bibnamefont
			{Panigrahi}}, \bibinfo {author} {\bibfnamefont {M.}~\bibnamefont {Sk}},
		\bibinfo {author} {\bibfnamefont {S.}~\bibnamefont {Jana}}, \bibinfo {author}
		{\bibfnamefont {S.}~\bibnamefont {Ghosh}}, \bibinfo {author} {\bibfnamefont
			{J.}~\bibnamefont {Deuermeier}}, \bibinfo {author} {\bibfnamefont
			{R.}~\bibnamefont {Martins}},\ and\ \bibinfo {author} {\bibfnamefont
			{E.}~\bibnamefont {Fortunato}},\ }\bibfield  {title} {\bibinfo {title}
		{{Tailoring the interface in high performance planar perovskite solar cell by
				ZnOS thin film}},\ }\href@noop {} {\bibfield  {journal} {\bibinfo  {journal}
			{ACS Appl. Energy Mater.}\ }\textbf {\bibinfo {volume} {5}},\ \bibinfo
		{pages} {5680} (\bibinfo {year} {2022})}\BibitemShut {NoStop}%
	\bibitem [{\citenamefont {Du}\ \emph {et~al.}(2020)\citenamefont {Du},
		\citenamefont {Wei}, \citenamefont {Cai}, \citenamefont {Liu}, \citenamefont
		{Wu}, \citenamefont {Li}, \citenamefont {Chen}, \citenamefont {Xia},
		\citenamefont {Xing},\ and\ \citenamefont {Huang}}]{du2020crystal}%
	\BibitemOpen
	\bibfield  {author} {\bibinfo {author} {\bibfnamefont {B.}~\bibnamefont
			{Du}}, \bibinfo {author} {\bibfnamefont {Q.}~\bibnamefont {Wei}}, \bibinfo
		{author} {\bibfnamefont {Y.}~\bibnamefont {Cai}}, \bibinfo {author}
		{\bibfnamefont {T.}~\bibnamefont {Liu}}, \bibinfo {author} {\bibfnamefont
			{B.}~\bibnamefont {Wu}}, \bibinfo {author} {\bibfnamefont {Y.}~\bibnamefont
			{Li}}, \bibinfo {author} {\bibfnamefont {Y.}~\bibnamefont {Chen}}, \bibinfo
		{author} {\bibfnamefont {Y.}~\bibnamefont {Xia}}, \bibinfo {author}
		{\bibfnamefont {G.}~\bibnamefont {Xing}},\ and\ \bibinfo {author}
		{\bibfnamefont {W.}~\bibnamefont {Huang}},\ }\bibfield  {title} {\bibinfo
		{title} {{Crystal face dependent charge carrier extraction in
				TiO$_2$/perovskite heterojunctions}},\ }\href@noop {} {\bibfield  {journal}
		{\bibinfo  {journal} {Nano Energy}\ }\textbf {\bibinfo {volume} {67}},\
		\bibinfo {pages} {104227} (\bibinfo {year} {2020})}\BibitemShut {NoStop}%
	\bibitem [{\citenamefont {Chen}\ \emph {et~al.}(2019)\citenamefont {Chen},
		\citenamefont {Messing}, \citenamefont {Zheng},\ and\ \citenamefont
		{Pullerits}}]{chen2019cation}%
	\BibitemOpen
	\bibfield  {author} {\bibinfo {author} {\bibfnamefont {J.}~\bibnamefont
			{Chen}}, \bibinfo {author} {\bibfnamefont {M.~E.}\ \bibnamefont {Messing}},
		\bibinfo {author} {\bibfnamefont {K.}~\bibnamefont {Zheng}},\ and\ \bibinfo
		{author} {\bibfnamefont {T.}~\bibnamefont {Pullerits}},\ }\bibfield  {title}
	{\bibinfo {title} {Cation-dependent hot carrier cooling in halide perovskite
			nanocrystals},\ }\href@noop {} {\bibfield  {journal} {\bibinfo  {journal} {J.
				Am. Chem. Soc.}\ }\textbf {\bibinfo {volume} {141}},\ \bibinfo {pages} {3532}
		(\bibinfo {year} {2019})}\BibitemShut {NoStop}%
	\bibitem [{\citenamefont {de~Weerd}\ \emph {et~al.}(2018)\citenamefont
		{de~Weerd}, \citenamefont {Gomez}, \citenamefont {Capretti}, \citenamefont
		{Lebrun}, \citenamefont {Matsubara}, \citenamefont {Lin}, \citenamefont
		{Ashida}, \citenamefont {Spoor}, \citenamefont {Siebbeles}, \citenamefont
		{Houtepen} \emph {et~al.}}]{de2018efficient}%
	\BibitemOpen
	\bibfield  {author} {\bibinfo {author} {\bibfnamefont {C.}~\bibnamefont
			{de~Weerd}}, \bibinfo {author} {\bibfnamefont {L.}~\bibnamefont {Gomez}},
		\bibinfo {author} {\bibfnamefont {A.}~\bibnamefont {Capretti}}, \bibinfo
		{author} {\bibfnamefont {D.~M.}\ \bibnamefont {Lebrun}}, \bibinfo {author}
		{\bibfnamefont {E.}~\bibnamefont {Matsubara}}, \bibinfo {author}
		{\bibfnamefont {J.}~\bibnamefont {Lin}}, \bibinfo {author} {\bibfnamefont
			{M.}~\bibnamefont {Ashida}}, \bibinfo {author} {\bibfnamefont {F.~C.}\
			\bibnamefont {Spoor}}, \bibinfo {author} {\bibfnamefont {L.~D.}\ \bibnamefont
			{Siebbeles}}, \bibinfo {author} {\bibfnamefont {A.~J.}\ \bibnamefont
			{Houtepen}}, \emph {et~al.},\ }\bibfield  {title} {\bibinfo {title}
		{{Efficient carrier multiplication in CsPbI$_3$ perovskite nanocrystals}},\
	}\href@noop {} {\bibfield  {journal} {\bibinfo  {journal} {Nat. Commun.}\
		}\textbf {\bibinfo {volume} {9}},\ \bibinfo {pages} {4199} (\bibinfo {year}
		{2018})}\BibitemShut {NoStop}%
	\bibitem [{\citenamefont {Nayak}\ and\ \citenamefont
		{Ghosh}(2025)}]{nayak2025optimizing}%
	\BibitemOpen
	\bibfield  {author} {\bibinfo {author} {\bibfnamefont {P.~K.}\ \bibnamefont
			{Nayak}}\ and\ \bibinfo {author} {\bibfnamefont {D.}~\bibnamefont {Ghosh}},\
	}\bibfield  {title} {\bibinfo {title} {Optimizing excited charge dynamics in
			layered halide perovskites through compositional engineering},\ }\href@noop
	{} {\bibfield  {journal} {\bibinfo  {journal} {Nano Lett.}\ } (\bibinfo
		{year} {2025})}\BibitemShut {NoStop}%
	\bibitem [{\citenamefont {Runge}\ and\ \citenamefont
		{Gross}(1984)}]{runge1984density}%
	\BibitemOpen
	\bibfield  {author} {\bibinfo {author} {\bibfnamefont {E.}~\bibnamefont
			{Runge}}\ and\ \bibinfo {author} {\bibfnamefont {E.~K.}\ \bibnamefont
			{Gross}},\ }\bibfield  {title} {\bibinfo {title} {Density-functional theory
			for time-dependent systems},\ }\href@noop {} {\bibfield  {journal} {\bibinfo
			{journal} {Phys. Rev. Lett.}\ }\textbf {\bibinfo {volume} {52}},\ \bibinfo
		{pages} {997} (\bibinfo {year} {1984})}\BibitemShut {NoStop}%
	\bibitem [{\citenamefont {Craig}\ \emph {et~al.}(2005)\citenamefont {Craig},
		\citenamefont {Duncan},\ and\ \citenamefont {Prezhdo}}]{craig2005trajectory}%
	\BibitemOpen
	\bibfield  {author} {\bibinfo {author} {\bibfnamefont {C.~F.}\ \bibnamefont
			{Craig}}, \bibinfo {author} {\bibfnamefont {W.~R.}\ \bibnamefont {Duncan}},\
		and\ \bibinfo {author} {\bibfnamefont {O.~V.}\ \bibnamefont {Prezhdo}},\
	}\bibfield  {title} {\bibinfo {title} {{Trajectory Surface Hopping in the
				Time-Dependent Kohn-Sham Approach for Electron-Nuclear Dynamics}},\
	}\href@noop {} {\bibfield  {journal} {\bibinfo  {journal} {Phys. Rev. Lett.}\
		}\textbf {\bibinfo {volume} {95}},\ \bibinfo {pages} {163001} (\bibinfo
		{year} {2005})}\BibitemShut {NoStop}%
	\bibitem [{\citenamefont {Jaeger}\ \emph {et~al.}(2012)\citenamefont {Jaeger},
		\citenamefont {Fischer},\ and\ \citenamefont
		{Prezhdo}}]{jaeger2012decoherence}%
	\BibitemOpen
	\bibfield  {author} {\bibinfo {author} {\bibfnamefont {H.~M.}\ \bibnamefont
			{Jaeger}}, \bibinfo {author} {\bibfnamefont {S.}~\bibnamefont {Fischer}},\
		and\ \bibinfo {author} {\bibfnamefont {O.~V.}\ \bibnamefont {Prezhdo}},\
	}\bibfield  {title} {\bibinfo {title} {Decoherence-induced surface hopping},\
	}\href@noop {} {\bibfield  {journal} {\bibinfo  {journal} {J. Chem. Phys.}\
		}\textbf {\bibinfo {volume} {137}} (\bibinfo {year} {2012})}\BibitemShut
	{NoStop}%
	\bibitem [{\citenamefont {Akimov}\ and\ \citenamefont
		{Prezhdo}(2013)}]{akimov2013pyxaid}%
	\BibitemOpen
	\bibfield  {author} {\bibinfo {author} {\bibfnamefont {A.~V.}\ \bibnamefont
			{Akimov}}\ and\ \bibinfo {author} {\bibfnamefont {O.~V.}\ \bibnamefont
			{Prezhdo}},\ }\bibfield  {title} {\bibinfo {title} {The pyxaid program for
			non-adiabatic molecular dynamics in condensed matter systems},\ }\href@noop
	{} {\bibfield  {journal} {\bibinfo  {journal} {J. Chem. Theory Comput.}\
		}\textbf {\bibinfo {volume} {9}},\ \bibinfo {pages} {4959} (\bibinfo {year}
		{2013})}\BibitemShut {NoStop}%
	\bibitem [{\citenamefont {Akimov}\ and\ \citenamefont
		{Prezhdo}(2014)}]{akimov2014advanced}%
	\BibitemOpen
	\bibfield  {author} {\bibinfo {author} {\bibfnamefont {A.~V.}\ \bibnamefont
			{Akimov}}\ and\ \bibinfo {author} {\bibfnamefont {O.~V.}\ \bibnamefont
			{Prezhdo}},\ }\bibfield  {title} {\bibinfo {title} {Advanced capabilities of
			the pyxaid program: integration schemes, decoherence effects, multiexcitonic
			states, and field-matter interaction},\ }\href@noop {} {\bibfield  {journal}
		{\bibinfo  {journal} {J. Chem. Theory Comput.}\ }\textbf {\bibinfo {volume}
			{10}},\ \bibinfo {pages} {789} (\bibinfo {year} {2014})}\BibitemShut
	{NoStop}%
	\bibitem [{\citenamefont {Hamm}(2005)}]{hamm2005principles}%
	\BibitemOpen
	\bibfield  {author} {\bibinfo {author} {\bibfnamefont {P.}~\bibnamefont
			{Hamm}},\ }\bibfield  {title} {\bibinfo {title} {Principles of nonlinear
			optical spectroscopy: A practical approach or: Mukamel for dummies},\
	}\href@noop {} {\bibfield  {journal} {\bibinfo  {journal} {University of
				Zurich}\ }\textbf {\bibinfo {volume} {41}},\ \bibinfo {pages} {77} (\bibinfo
		{year} {2005})}\BibitemShut {NoStop}%
	\bibitem [{\citenamefont {Hohenberg}\ and\ \citenamefont
		{Kohn}(1964)}]{hohenberg1964inhomogeneous}%
	\BibitemOpen
	\bibfield  {author} {\bibinfo {author} {\bibfnamefont {P.}~\bibnamefont
			{Hohenberg}}\ and\ \bibinfo {author} {\bibfnamefont {W.}~\bibnamefont
			{Kohn}},\ }\bibfield  {title} {\bibinfo {title} {Inhomogeneous electron
			gas},\ }\href@noop {} {\bibfield  {journal} {\bibinfo  {journal} {Phys.
				Rev.}\ }\textbf {\bibinfo {volume} {136}},\ \bibinfo {pages} {B864} (\bibinfo
		{year} {1964})}\BibitemShut {NoStop}%
	\bibitem [{\citenamefont {Kohn}\ and\ \citenamefont
		{Sham}(1965)}]{kohn1965self}%
	\BibitemOpen
	\bibfield  {author} {\bibinfo {author} {\bibfnamefont {W.}~\bibnamefont
			{Kohn}}\ and\ \bibinfo {author} {\bibfnamefont {L.~J.}\ \bibnamefont
			{Sham}},\ }\bibfield  {title} {\bibinfo {title} {Self-consistent equations
			including exchange and correlation effects},\ }\href@noop {} {\bibfield
		{journal} {\bibinfo  {journal} {Phys. Rev.}\ }\textbf {\bibinfo {volume}
			{140}},\ \bibinfo {pages} {A1133} (\bibinfo {year} {1965})}\BibitemShut
	{NoStop}%
	\bibitem [{\citenamefont {Kresse}\ and\ \citenamefont
		{Furthm{\"u}ller}(1996)}]{kresse1996efficiency}%
	\BibitemOpen
	\bibfield  {author} {\bibinfo {author} {\bibfnamefont {G.}~\bibnamefont
			{Kresse}}\ and\ \bibinfo {author} {\bibfnamefont {J.}~\bibnamefont
			{Furthm{\"u}ller}},\ }\bibfield  {title} {\bibinfo {title} {Efficiency of
			ab-initio total energy calculations for metals and semiconductors using a
			plane-wave basis set},\ }\href@noop {} {\bibfield  {journal} {\bibinfo
			{journal} {Comput. Mater. Sci.}\ }\textbf {\bibinfo {volume} {6}},\ \bibinfo
		{pages} {15} (\bibinfo {year} {1996})}\BibitemShut {NoStop}%
	\bibitem [{\citenamefont {Kresse}\ and\ \citenamefont
		{Joubert}(1999)}]{kresse1999ultrasoft}%
	\BibitemOpen
	\bibfield  {author} {\bibinfo {author} {\bibfnamefont {G.}~\bibnamefont
			{Kresse}}\ and\ \bibinfo {author} {\bibfnamefont {D.}~\bibnamefont
			{Joubert}},\ }\bibfield  {title} {\bibinfo {title} {From ultrasoft
			pseudopotentials to the projector augmented-wave method},\ }\href@noop {}
	{\bibfield  {journal} {\bibinfo  {journal} {Phys. Rev. B.}\ }\textbf
		{\bibinfo {volume} {59}},\ \bibinfo {pages} {1758} (\bibinfo {year}
		{1999})}\BibitemShut {NoStop}%
	\bibitem [{\citenamefont {Bl{\"o}chl}(1994)}]{blochl1994projector}%
	\BibitemOpen
	\bibfield  {author} {\bibinfo {author} {\bibfnamefont {P.~E.}\ \bibnamefont
			{Bl{\"o}chl}},\ }\bibfield  {title} {\bibinfo {title} {Projector
			augmented-wave method},\ }\href@noop {} {\bibfield  {journal} {\bibinfo
			{journal} {Phys. Rev. B.}\ }\textbf {\bibinfo {volume} {50}},\ \bibinfo
		{pages} {17953} (\bibinfo {year} {1994})}\BibitemShut {NoStop}%
	\bibitem [{\citenamefont {Perdew}\ \emph {et~al.}(1996)\citenamefont {Perdew},
		\citenamefont {Burke},\ and\ \citenamefont
		{Ernzerhof}}]{perdew1996generalized}%
	\BibitemOpen
	\bibfield  {author} {\bibinfo {author} {\bibfnamefont {J.~P.}\ \bibnamefont
			{Perdew}}, \bibinfo {author} {\bibfnamefont {K.}~\bibnamefont {Burke}},\ and\
		\bibinfo {author} {\bibfnamefont {M.}~\bibnamefont {Ernzerhof}},\ }\bibfield
	{title} {\bibinfo {title} {Generalized gradient approximation made simple},\
	}\href@noop {} {\bibfield  {journal} {\bibinfo  {journal} {Phys. Rev. Lett.}\
		}\textbf {\bibinfo {volume} {77}},\ \bibinfo {pages} {3865} (\bibinfo {year}
		{1996})}\BibitemShut {NoStop}%
		\bibitem{Supporting_Information}
		See Supporting Information at [URL] for the optimized lattice parameters; band gaps using $\langle$PBE$\rangle$, HSE06 (without and with SOC), $G_0W_0$@HSE06+SOC, and experimental reported values; variation of band gap with time, for $\mathrm{Ca_3NSb}$, $\mathrm{Sr_3NSb}$, $\mathrm{Sr_3NSb_{hexa}}$, and $\mathrm{Ba_3NSb_{hexa}}$. The Supporting Information also contains Refs. [16, 17].
	\bibitem [{\citenamefont {Stoiber}\ and\ \citenamefont
		{Niewa}(2019)}]{stoiber2019perovskite}%
	\BibitemOpen
	\bibfield  {author} {\bibinfo {author} {\bibfnamefont {D.}~\bibnamefont
			{Stoiber}}\ and\ \bibinfo {author} {\bibfnamefont {R.}~\bibnamefont
			{Niewa}},\ }\bibfield  {title} {\bibinfo {title} {Perovskite distortion
			inverted: Crystal structures of $\mathrm{(A_3N) As}$ ($\mathrm{A= Mg, Ca, Sr,
				Ba}$)},\ }\href@noop {} {\bibfield  {journal} {\bibinfo  {journal} {Z Anorg
				Allg Chem}\ }\textbf {\bibinfo {volume} {645}},\ \bibinfo {pages} {329}
		(\bibinfo {year} {2019})}\BibitemShut {NoStop}%
	\bibitem [{\citenamefont {Krukau}\ \emph {et~al.}(2006)\citenamefont {Krukau},
		\citenamefont {Vydrov}, \citenamefont {Izmaylov},\ and\ \citenamefont
		{Scuseria}}]{krukau2006influence}%
	\BibitemOpen
	\bibfield  {author} {\bibinfo {author} {\bibfnamefont {A.~V.}\ \bibnamefont
			{Krukau}}, \bibinfo {author} {\bibfnamefont {O.~A.}\ \bibnamefont {Vydrov}},
		\bibinfo {author} {\bibfnamefont {A.~F.}\ \bibnamefont {Izmaylov}},\ and\
		\bibinfo {author} {\bibfnamefont {G.~E.}\ \bibnamefont {Scuseria}},\
	}\bibfield  {title} {\bibinfo {title} {Influence of the exchange screening
			parameter on the performance of screened hybrid functionals},\ }\href@noop {}
	{\bibfield  {journal} {\bibinfo  {journal} {J. Chem. Phys.}\ }\textbf
		{\bibinfo {volume} {125}},\ \bibinfo {pages} {224106} (\bibinfo {year}
		{2006})}\BibitemShut {NoStop}%
	\bibitem [{\citenamefont {Giannozzi}\ \emph {et~al.}(2009)\citenamefont
		{Giannozzi}, \citenamefont {Baroni}, \citenamefont {Bonini}, \citenamefont
		{Calandra}, \citenamefont {Car}, \citenamefont {Cavazzoni}, \citenamefont
		{Ceresoli}, \citenamefont {Chiarotti}, \citenamefont {Cococcioni},
		\citenamefont {Dabo} \emph {et~al.}}]{giannozzi2009quantum}%
	\BibitemOpen
	\bibfield  {author} {\bibinfo {author} {\bibfnamefont {P.}~\bibnamefont
			{Giannozzi}}, \bibinfo {author} {\bibfnamefont {S.}~\bibnamefont {Baroni}},
		\bibinfo {author} {\bibfnamefont {N.}~\bibnamefont {Bonini}}, \bibinfo
		{author} {\bibfnamefont {M.}~\bibnamefont {Calandra}}, \bibinfo {author}
		{\bibfnamefont {R.}~\bibnamefont {Car}}, \bibinfo {author} {\bibfnamefont
			{C.}~\bibnamefont {Cavazzoni}}, \bibinfo {author} {\bibfnamefont
			{D.}~\bibnamefont {Ceresoli}}, \bibinfo {author} {\bibfnamefont {G.~L.}\
			\bibnamefont {Chiarotti}}, \bibinfo {author} {\bibfnamefont {M.}~\bibnamefont
			{Cococcioni}}, \bibinfo {author} {\bibfnamefont {I.}~\bibnamefont {Dabo}},
		\emph {et~al.},\ }\bibfield  {title} {\bibinfo {title} {{QUANTUM ESPRESSO: a
				modular and open-source software project for quantum simulations of
				materials}},\ }\href@noop {} {\bibfield  {journal} {\bibinfo  {journal} {J.
				Phys.: Condens. Matter}\ }\textbf {\bibinfo {volume} {21}},\ \bibinfo {pages}
		{395502} (\bibinfo {year} {2009})}\BibitemShut {NoStop}%
	\bibitem [{\citenamefont {Wang}\ \emph
		{et~al.}(2025{\natexlab{b}})\citenamefont {Wang}, \citenamefont {Chu},
		\citenamefont {Wu}, \citenamefont {Saidi},\ and\ \citenamefont
		{Prezhdo}}]{wang2025sub}%
	\BibitemOpen
	\bibfield  {author} {\bibinfo {author} {\bibfnamefont {B.}~\bibnamefont
			{Wang}}, \bibinfo {author} {\bibfnamefont {W.}~\bibnamefont {Chu}}, \bibinfo
		{author} {\bibfnamefont {Y.}~\bibnamefont {Wu}}, \bibinfo {author}
		{\bibfnamefont {W.~A.}\ \bibnamefont {Saidi}},\ and\ \bibinfo {author}
		{\bibfnamefont {O.~V.}\ \bibnamefont {Prezhdo}},\ }\bibfield  {title}
	{\bibinfo {title} {Sub-bandgap charge harvesting and energy up-conversion in
			metal halide perovskites: ab initio quantum dynamics},\ }\href@noop {}
	{\bibfield  {journal} {\bibinfo  {journal} {npj Comput. Mater.}\ }\textbf
		{\bibinfo {volume} {11}},\ \bibinfo {pages} {11} (\bibinfo {year}
		{2025}{\natexlab{b}})}\BibitemShut {NoStop}%
	\bibitem [{\citenamefont {Yang}\ \emph {et~al.}(2025)\citenamefont {Yang},
		\citenamefont {Cai}, \citenamefont {Xia}, \citenamefont {Liu}, \citenamefont
		{Ma}, \citenamefont {Zhang}, \citenamefont {Liu}, \citenamefont {Cao},
		\citenamefont {Shen}, \citenamefont {Chen} \emph
		{et~al.}}]{yang2025reducing}%
	\BibitemOpen
	\bibfield  {author} {\bibinfo {author} {\bibfnamefont {B.}~\bibnamefont
			{Yang}}, \bibinfo {author} {\bibfnamefont {B.}~\bibnamefont {Cai}}, \bibinfo
		{author} {\bibfnamefont {J.}~\bibnamefont {Xia}}, \bibinfo {author}
		{\bibfnamefont {Y.}~\bibnamefont {Liu}}, \bibinfo {author} {\bibfnamefont
			{Y.}~\bibnamefont {Ma}}, \bibinfo {author} {\bibfnamefont {J.}~\bibnamefont
			{Zhang}}, \bibinfo {author} {\bibfnamefont {L.}~\bibnamefont {Liu}}, \bibinfo
		{author} {\bibfnamefont {K.}~\bibnamefont {Cao}}, \bibinfo {author}
		{\bibfnamefont {W.}~\bibnamefont {Shen}}, \bibinfo {author} {\bibfnamefont
			{S.}~\bibnamefont {Chen}}, \emph {et~al.},\ }\bibfield  {title} {\bibinfo
		{title} {Reducing nonradiative recombination in halide perovskites through
			appropriate band gaps and heavy atomic masses},\ }\href@noop {} {\bibfield
		{journal} {\bibinfo  {journal} {J. Phys. Chem. Lett.}\ }\textbf {\bibinfo
			{volume} {16}},\ \bibinfo {pages} {1253} (\bibinfo {year}
		{2025})}\BibitemShut {NoStop}%
	\bibitem [{\citenamefont {Prasanna}\ \emph {et~al.}(2017)\citenamefont
		{Prasanna}, \citenamefont {Gold-Parker}, \citenamefont {Leijtens},
		\citenamefont {Conings}, \citenamefont {Babayigit}, \citenamefont {Boyen},
		\citenamefont {Toney},\ and\ \citenamefont {McGehee}}]{prasanna2017band}%
	\BibitemOpen
	\bibfield  {author} {\bibinfo {author} {\bibfnamefont {R.}~\bibnamefont
			{Prasanna}}, \bibinfo {author} {\bibfnamefont {A.}~\bibnamefont
			{Gold-Parker}}, \bibinfo {author} {\bibfnamefont {T.}~\bibnamefont
			{Leijtens}}, \bibinfo {author} {\bibfnamefont {B.}~\bibnamefont {Conings}},
		\bibinfo {author} {\bibfnamefont {A.}~\bibnamefont {Babayigit}}, \bibinfo
		{author} {\bibfnamefont {H.-G.}\ \bibnamefont {Boyen}}, \bibinfo {author}
		{\bibfnamefont {M.~F.}\ \bibnamefont {Toney}},\ and\ \bibinfo {author}
		{\bibfnamefont {M.~D.}\ \bibnamefont {McGehee}},\ }\bibfield  {title}
	{\bibinfo {title} {Band gap tuning via lattice contraction and octahedral
			tilting in perovskite materials for photovoltaics},\ }\href@noop {}
	{\bibfield  {journal} {\bibinfo  {journal} {J. Am. Chem. Soc.}\ }\textbf
		{\bibinfo {volume} {139}},\ \bibinfo {pages} {11117} (\bibinfo {year}
		{2017})}\BibitemShut {NoStop}%
	\bibitem [{\citenamefont {Knutson}\ \emph {et~al.}(2005)\citenamefont
		{Knutson}, \citenamefont {Martin},\ and\ \citenamefont
		{Mitzi}}]{knutson2005tuning}%
	\BibitemOpen
	\bibfield  {author} {\bibinfo {author} {\bibfnamefont {J.~L.}\ \bibnamefont
			{Knutson}}, \bibinfo {author} {\bibfnamefont {J.~D.}\ \bibnamefont
			{Martin}},\ and\ \bibinfo {author} {\bibfnamefont {D.~B.}\ \bibnamefont
			{Mitzi}},\ }\bibfield  {title} {\bibinfo {title} {Tuning the band gap in
			hybrid tin iodide perovskite semiconductors using structural templating},\
	}\href@noop {} {\bibfield  {journal} {\bibinfo  {journal} {Inorg. Chem.}\
		}\textbf {\bibinfo {volume} {44}},\ \bibinfo {pages} {4699} (\bibinfo {year}
		{2005})}\BibitemShut {NoStop}%
	\bibitem [{\citenamefont {Kong}\ \emph {et~al.}(2016)\citenamefont {Kong},
		\citenamefont {Liu}, \citenamefont {Gong}, \citenamefont {Hu}, \citenamefont
		{Schaller}, \citenamefont {Dera}, \citenamefont {Zhang}, \citenamefont {Liu},
		\citenamefont {Yang}, \citenamefont {Zhu} \emph
		{et~al.}}]{kong2016simultaneous}%
	\BibitemOpen
	\bibfield  {author} {\bibinfo {author} {\bibfnamefont {L.}~\bibnamefont
			{Kong}}, \bibinfo {author} {\bibfnamefont {G.}~\bibnamefont {Liu}}, \bibinfo
		{author} {\bibfnamefont {J.}~\bibnamefont {Gong}}, \bibinfo {author}
		{\bibfnamefont {Q.}~\bibnamefont {Hu}}, \bibinfo {author} {\bibfnamefont
			{R.~D.}\ \bibnamefont {Schaller}}, \bibinfo {author} {\bibfnamefont
			{P.}~\bibnamefont {Dera}}, \bibinfo {author} {\bibfnamefont {D.}~\bibnamefont
			{Zhang}}, \bibinfo {author} {\bibfnamefont {Z.}~\bibnamefont {Liu}}, \bibinfo
		{author} {\bibfnamefont {W.}~\bibnamefont {Yang}}, \bibinfo {author}
		{\bibfnamefont {K.}~\bibnamefont {Zhu}}, \emph {et~al.},\ }\bibfield  {title}
	{\bibinfo {title} {Simultaneous band-gap narrowing and carrier-lifetime
			prolongation of organic--inorganic trihalide perovskites},\ }\href@noop {}
	{\bibfield  {journal} {\bibinfo  {journal} {Proc. Natl. Acad. Sci. U.S.A.}\
		}\textbf {\bibinfo {volume} {113}},\ \bibinfo {pages} {8910} (\bibinfo {year}
		{2016})}\BibitemShut {NoStop}%
	\bibitem [{\citenamefont {Prezhdo}(2000)}]{prezhdo2000quantum}%
	\BibitemOpen
	\bibfield  {author} {\bibinfo {author} {\bibfnamefont {O.~V.}\ \bibnamefont
			{Prezhdo}},\ }\bibfield  {title} {\bibinfo {title} {Quantum anti-zeno
			acceleration of a chemical reaction},\ }\href@noop {} {\bibfield  {journal}
		{\bibinfo  {journal} {Phys. Rev. Lett.}\ }\textbf {\bibinfo {volume} {85}},\
		\bibinfo {pages} {4413} (\bibinfo {year} {2000})}\BibitemShut {NoStop}%
	\bibitem [{\citenamefont {Kilina}\ \emph {et~al.}(2013)\citenamefont {Kilina},
		\citenamefont {Neukirch}, \citenamefont {Habenicht}, \citenamefont {Kilin},\
		and\ \citenamefont {Prezhdo}}]{kilina2013quantum}%
	\BibitemOpen
	\bibfield  {author} {\bibinfo {author} {\bibfnamefont {S.~V.}\ \bibnamefont
			{Kilina}}, \bibinfo {author} {\bibfnamefont {A.~J.}\ \bibnamefont
			{Neukirch}}, \bibinfo {author} {\bibfnamefont {B.~F.}\ \bibnamefont
			{Habenicht}}, \bibinfo {author} {\bibfnamefont {D.~S.}\ \bibnamefont
			{Kilin}},\ and\ \bibinfo {author} {\bibfnamefont {O.~V.}\ \bibnamefont
			{Prezhdo}},\ }\bibfield  {title} {\bibinfo {title} {Quantum zeno effect
			rationalizes the phonon bottleneck in semiconductor quantum dots},\
	}\href@noop {} {\bibfield  {journal} {\bibinfo  {journal} {Phys. Rev. Lett.}\
		}\textbf {\bibinfo {volume} {110}},\ \bibinfo {pages} {180404} (\bibinfo
		{year} {2013})}\BibitemShut {NoStop}%
\end{thebibliography}
%
\end{document}